\documentclass[twocolumn,prb,superscriptaddress,showpacs,floats]{revtex4}

\usepackage{graphicx}
\usepackage{times}
\usepackage{amsmath,amssymb}

\begin{document}

\title{Quantum spin ladders of non-Abelian anyons}

\author{Didier Poilblanc}
\affiliation{Laboratoire de Physique Th\'eorique, CNRS and Universit\'e 
de Toulouse, F-31062 Toulouse, France}
\author{Andreas W.W. Ludwig}
\affiliation{Physics Department, University of California, Santa Barbara, California 93106}
\author{Simon Trebst}
\affiliation{Microsoft Research, Station Q,
University of California, Santa Barbara, CA 93106} 
\author{Matthias Troyer}
\affiliation{Theoretische Physik, ETH Zurich, 8093 Zurich, Switzerland}

\date{\today}

\begin{abstract}

Quantum ladder models, consisting of coupled chains, form intriguing systems bridging one and 
two dimensions and have been well studied in the context of quantum magnets and fermionic systems. 
Here we consider ladder systems made of more exotic quantum mechanical degrees of freedom, so-called non-Abelian anyons, which can be thought of as certain quantum deformations of ordinary SU(2) spins. Such non-Abelian anyons occur as quasiparticle excitations in topological quantum fluids, including $p_x + i p_y$ superconductors, certain fractional quantum Hall states, and rotating Bose-Einstein condensates. 
Here we use a combination of exact diagonalization and conformal field theory to determine the phase diagrams of ladders with up to four chains.
We discuss how phenomena familiar from ordinary SU(2) spin ladders are generalized in their anyonic counterparts, such as gapless and gapped phases, odd/even effects with the ladder width, and elementary `magnon' excitations. Other features are entirely due to the topological nature of the anyonic degrees of freedom. In general, two-dimensional systems of interacting localized non-Abelian anyons are anyonic generalizations of two-dimensional quantum magnets.

\end{abstract}

\maketitle

\section{Introduction}

Quantum antiferromagnets and, more generally, electronic systems are notoriously known to behave fundamentally differently 
in one and two spatial dimensions. In one spatial dimension quantum fluctuations are enhanced and often give rise to critical 
properties such as algebraic  spin (or charge) correlations~\cite{1D}. In contrast, in two spatial dimensions one frequently finds 
quantum ground states with long-range order, which often originate from the spontaneous breaking of a continuous symmetry.
The archetypal example of the latter are Heisenberg antiferromagnets on bipartite lattices~\cite{2D_Heisenberg}, where the N\'eel 
ground state arises from the spontaneous breaking of the SU(2) spin symmetry and the resulting zero-energy Goldstrone modes 
are spin-wave excitations, also called magnons.
Quantum ladder systems, consisting of a finite number $W$ of coupled one-dimensional (1D) systems, form a bridge between 
between these two limits, and the evolution of quantum ground states in the dimensional crossover of increasing ladder width
has been well studied in the context of itinerant bosonic and fermionic systems \cite{ItinerantLadders} 
as well as quantum spin ladders \cite{OrdinaryLadders}. 
A variety of remarkable crossover effects have been observed, such as the celebrated even/odd effect in quantum spin-1/2
Heisenberg ladders, where gapless ground states are found for all odd width $W$, while ladder systems with an even number of legs
exhibit a spin gap \cite{OrdinaryLadders}. Surely, this effect has also been experimentally observed for actual materials realizing almost 
perfect two and three leg S=1/2 antiferromagnetic (AFM) ladders when performing careful magnetic susceptibility measurements
\cite{ExpLadders}.

In this manuscript, we consider systems of more exotic quantum mechanical degrees of freedom, so-called non-Abelian anyons,
which have attracted considerable interest in the description of non-Abelian vortices in unconventional $p_x + i p_y$ 
superconductors~\cite{ReadGreen}, quasiholes in certain fractional quantum Hall states~\cite{MooreRead,Kareljan09,Bonderson08}, or vortices in rotating
Bose-Einstein condensates~\cite{Cooper01}, and in theoretical proposals for inherently fault tolerant quantum computing schemes 
\cite{Nayak08}.
Since we are interested in their collective quantum ground states in two spatial dimensions, we follow a route similar to the 
above-mentioned studies of SU(2) quantum antiferromagnets and study ladder systems of interacting anyonic degrees of freedom. 

Formally, non-Abelian anyons can be described by so-called su(2)$_k$ Chern-Simons theories, which correspond to certain
quantum deformations \cite{SU2q} of SU(2). In these theories, the non-Abelian degrees of freedoms are captured by `generalized
angular momenta' $j$, which for a given su(2)$_k$ theory, are constrained to  the first $k+1$ representations of SU(2)
\[
    j = 0, \frac{1}{2}, 1, \frac{3}{2}, \ldots, \frac{k}{2} \,.
\]
Similar to the coupling of ordinary angular momenta, two non-Abelian degrees
of freedom can be `fused' into multiple states with total angular momenta (or spins)
\[
   j_1 \otimes j_2 = |j_1-j_2| \oplus |j_1-j_2|+1 \oplus \ldots \oplus \min(j_1+j_2, k-j_1-j_2) \,,
\]
where again the `cutoff' $k$ of the deformation enters.
The occurrence of {\sl multiple} fusion channels on the right-hand-side of the above equation is what intrinsically gives rise 
to a {\sl macroscopic} degeneracy of states for a set of multiple non-Abelian anyons -- the hallmark of non-Abelian statistics.

In this manuscript we consider the fundamental case of non-Abelian anyons with generalized angular momentum $j=1/2$, which obey the fusion rule $1/2 \otimes 1/2 = 0 \oplus 1 $ reminiscent of two ordinary spin-1/2's coupling into a singlet or triplet.
Like the Heisenberg Hamiltonian for ordinary spins, 
interactions between the anyons energetically split the two fusion outcomes, which in the case
of ordinary SU(2) spins is captured by the familiar Heisenberg Hamiltonian
\begin{eqnarray}
    H & = & \sum_{i,j} J_{ij} \vec{S}_i \cdot \vec{S}_j \\
                                     & = & \frac{1}{2} \sum_{i,j} J_{ij} \left[ (\vec{S}_i+\vec{S}_j)^2 - \vec{S}_i^2 - \vec{S}_j^2 \right] \\
                                     & = & -\sum_{i,j} J_{ij} \Pi_{i,j}^0 + {\rm const.} \,, \label{Eq:Projector}
\end{eqnarray}
which can be viewed as a sum of pairwise projectors $\Pi_{i,j}^0$ onto the singlet state.
Similarly, we can define an `anyonic Heisenberg Hamiltonian' that for a pair of non-Abelian anyons with generalized angular momentum $j=1/2$ projects onto the $j=0$ (singlet) fusion channel, thus taking the same form as Eq.~\eqref{Eq:Projector} of the Hamiltonian above. 
In analogy to ordinary SU(2) spins we refer to positive couplings (projecting onto the generalized $j=0$
state) as `antiferromagnetic', while negative couplings (projecting onto the generalized $j=1$ state)
are `ferromagnetic'. 

\begin{figure}[b]
\begin{center}
	\includegraphics[width=0.9\columnwidth]{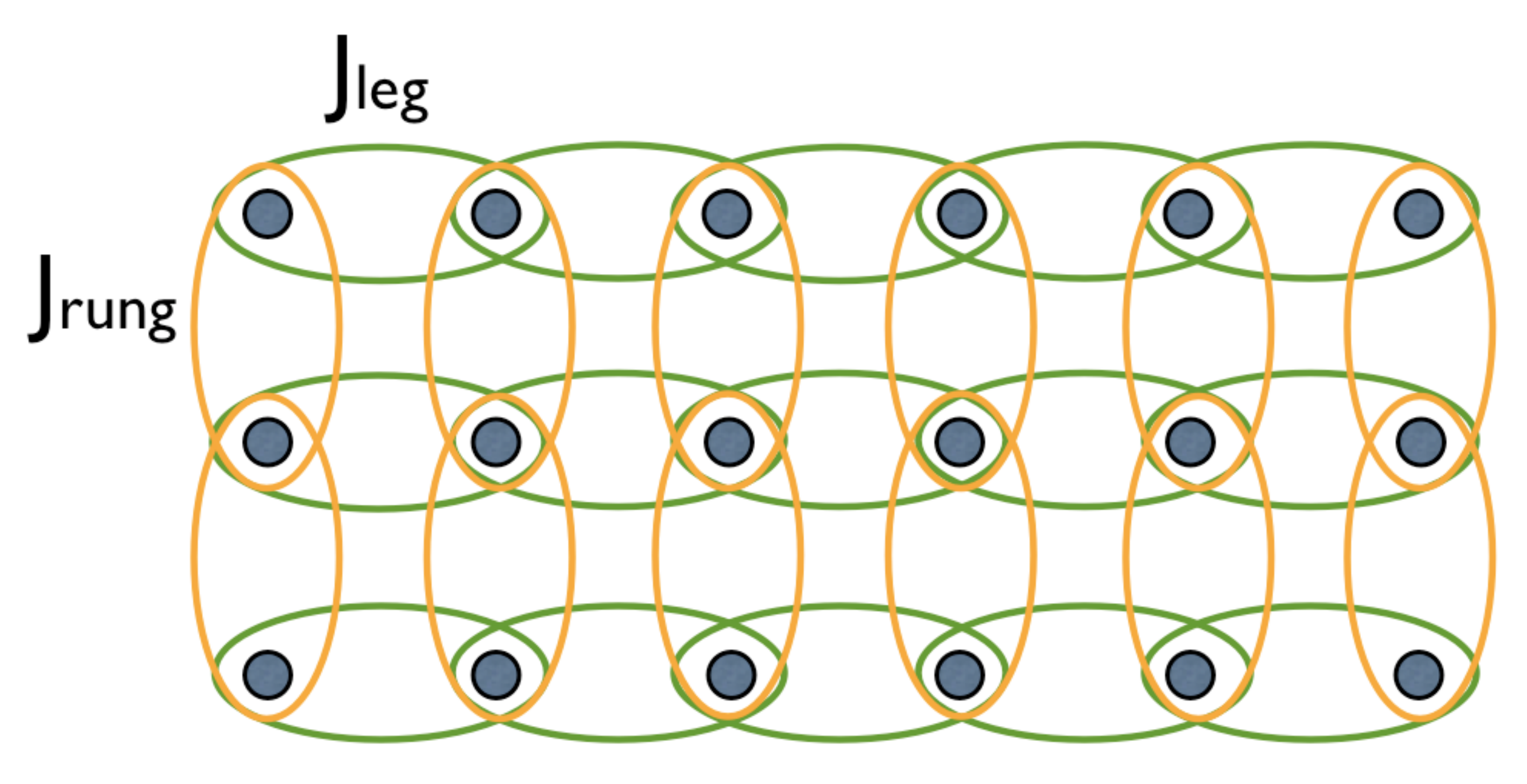}
\end{center}
\caption{
   Three coupled chains of interacting anyons (indicated by the filled circles).
   The interaction along ($J_{\rm leg}$) and perpendicular ($J_{\rm rung}$) to the chains are indicated by the ellipses.
} 
\label{Fig:ThreeLeg} 
\end{figure}

It has recently been shown that similar to their SU(2) counterparts chains of interacting non-Abelian anyons can exhibit a variety of collective ground states including stable gapless phases
\cite{Feiguin07,Trebst08,Gils09,Ardonne10}
and exotic infinite-randomness fixed points
\cite{Bonesteel07,Fidkowski08, Fidkowski09}.
In this manuscript, we aim at understanding {\sl two-dimensional} ground states of interacting non-Abelian
anyons and -- following a similar route as in the case of the above-mentioned studies of SU(2) quantum antiferromagnets -- we consider systems of coupled chains forming $W$-leg ladders.
Employing extensive numerical simulations combined with a conformal field theory analysis, we investigate phase diagrams of $W$-leg ladder with up to $W=4$ legs, which allows us to derive some conclusions also for the 2D limit of $W\rightarrow\infty$.
We mostly focus on the case of su(2)$_k$ with $k=3$ as a representative example and in particular all numerical simulations are performed for this case. We will return to the more general case of arbitrary level $k>3$ in Sec.~\ref{Sec:SU2k}.

While in this manuscript we detail the physics of interacting non-Abelian anyons mostly in terms of (deformed) quantum spins -- a notion more familiar to the field of low-dimensional quantum magnetism,  we have put forward another perspective on the physics of interacting anyons in the context of certain fractional quantum Hall states in a recent article \cite{Ludwig10}.
There we have made a connection between the collective states of (anyonic) excitations in non-Abelian quantum Hall liquids and the physics of moving on a non-Abelian quantum Hall plateau.

The remainder of this manuscript is structured as follows: We will start with a detailed derivation
of the microscopic models analyzed in this manuscript in Sec.~\ref{Sec:Model}. This is followed
by a discussion of the phase diagrams of the various ladder models starting from the strong rung-coupling limit in Sec.~\ref{Sec:StrongCoupling} and continuing with the weak rung-coupling limit in Sec.~\ref{Sec:Liquids}. We will then turn to the peculiar role of boundary conditions and the occurrence of 
gapless modes at open boundaries for these anyonic ladder models in Sec.~\ref{Sec:Boundaries}.
We round off the manuscript by a  discussion of the two-dimensional limit of these ladder models in Sec.~\ref{Sec:2D} and generalization to su(2)$_k$ theories with $k > 3$ in Sec.~\ref{Sec:SU2k}.

\begin{figure}
\begin{center}
  \includegraphics*[width=0.35\textwidth,angle=0]{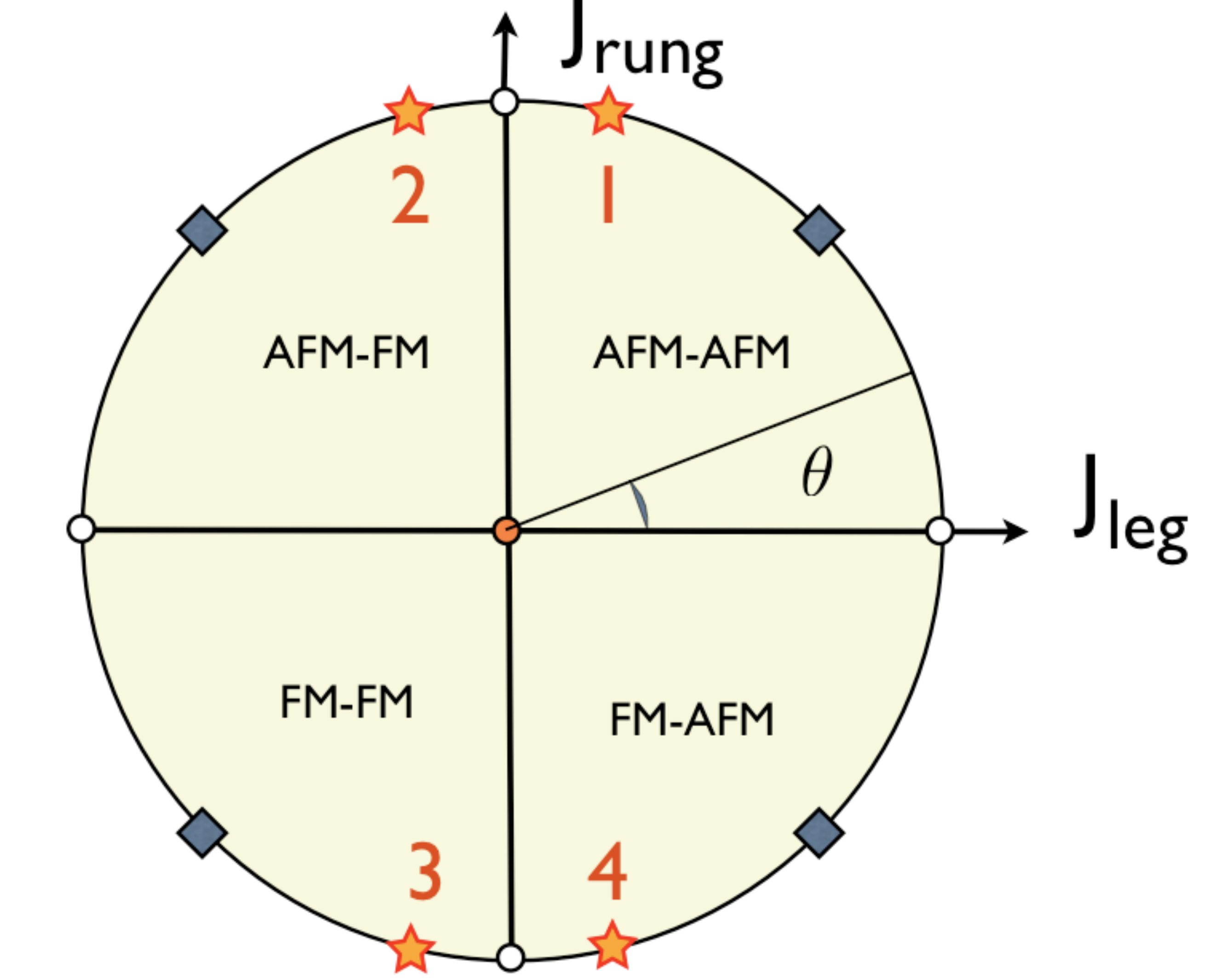}
\end{center}
\caption {
(color online)
Representation of the parameter space of the model on a circle. 
The couplings $J_{\rm rung}$ and $J_{\rm leg}$ can be either positive (AFM) or negative (FM).
The different phases in each of the quadrants labelled by "$J_{\rm rung}$"--"$J_{\rm leg}$" are given in Table~\ref{table:phase_diagram}.
The (grey) squares at $\pi/4$ correspond to isotropic couplings $|J_{\rm rung}/J_{\rm leg}|=1$.
The (orange) stars mark the parameters in the vicinity of the strong coupling limit 
$|J_{\rm rung}| \gg |J_{\rm leg}|$ 
used in this work, namely $\theta=\{ 3\pi/7, 4\pi/7, 10\pi/7, 11\pi/7 \}$ labeled from 1 to 4, respectively.
}
\label{Fig:Circle}
\end{figure}


\section{The microscopic ladder model}
\label{Sec:Model}

In this section we will give a definition of the microscopic $W$-leg ladder models for so-called 
su(2)$_3$ Fibonacci anyons. 
We will keep our discussion short but self-contained, as an extended derivation of general
microscopic Hamiltonians has been given in Ref.~\onlinecite{Proceedings}. We will emphasize 
in the following those aspects that are not covered in Ref.~\onlinecite{Proceedings}. 
For a given $W$-leg ladder we denote the strength of the interactions as  $J_{\rm leg}$ and $J_{\rm rung}$ for the coupling along and perpendicular to the chains, respectively, as illustrated in Fig.~\ref{Fig:ThreeLeg}.  Parametrizing these couplings as $J_{\rm leg} = \cos \theta$ and $J_{\rm rung} = \sin \theta$ we will map out the parameter space on a unit circle as shown in Fig.~\ref{Fig:Circle}.

Our numerical analysis of these ladder systems is based on exact diagonalization using the Lanczos algorithm which provides us with the low-energy spectra of finite systems with extent $W \times L$,
where $L$ is the length of the ladder in the chain direction and $W$ is the width of the ladder in the rung direction. In our exact diagonaliztion studies, we have been able to analyze systems of size $2\times L$ ($L=8,12,15,16,18,20,21$), $3\times L$ ($L=6,8,9,10,12$) and $4\times L$ ($L=4,6,8$). 

\subsection{The basis states}

\begin{figure}
\begin{center}
\includegraphics[width=\columnwidth]{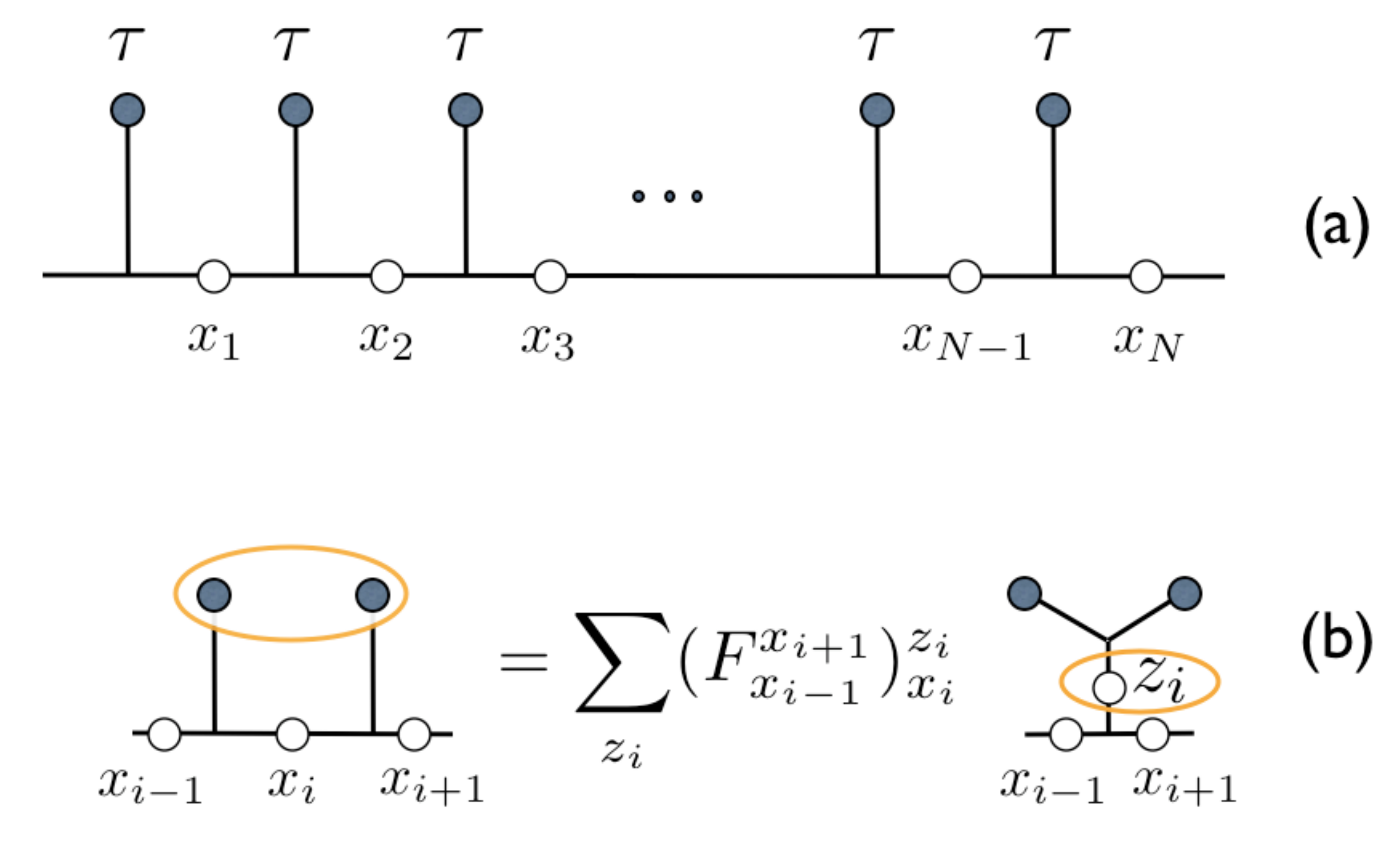}\end{center}
\begin{center}
\includegraphics[width=\columnwidth]{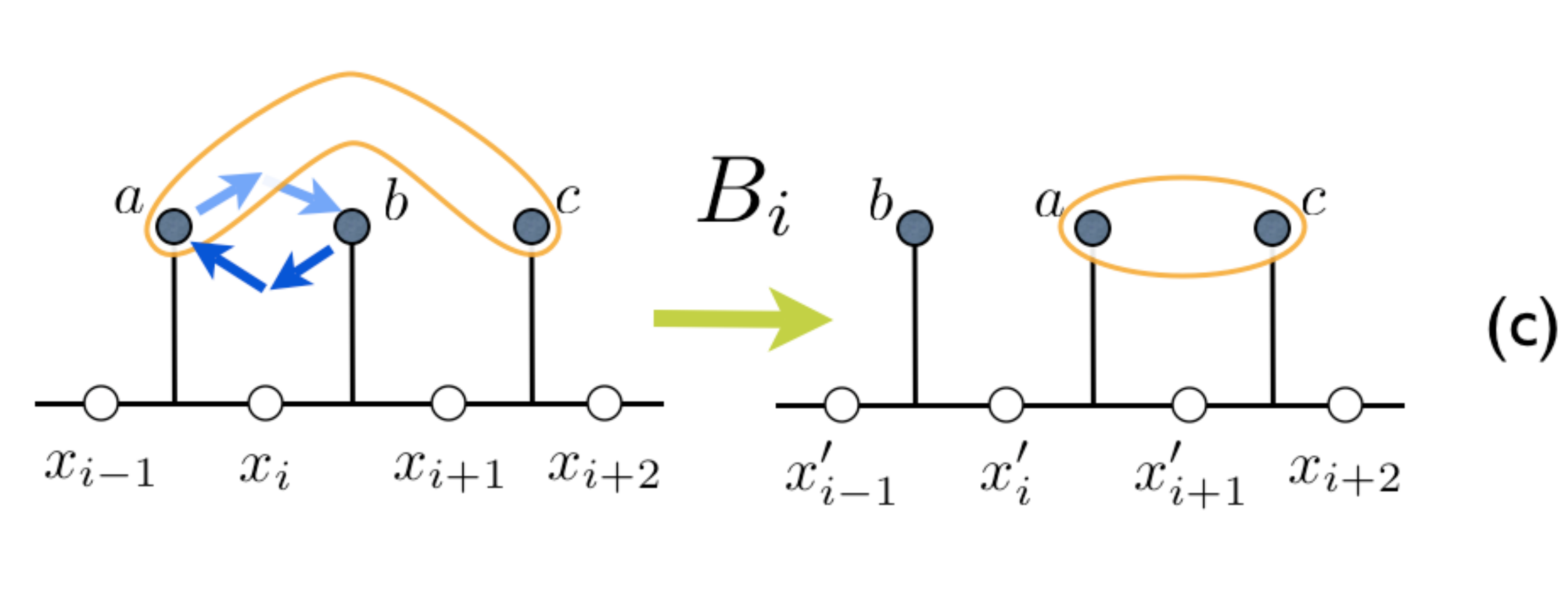}\end{center}
\begin{center}
\includegraphics[width=\columnwidth]{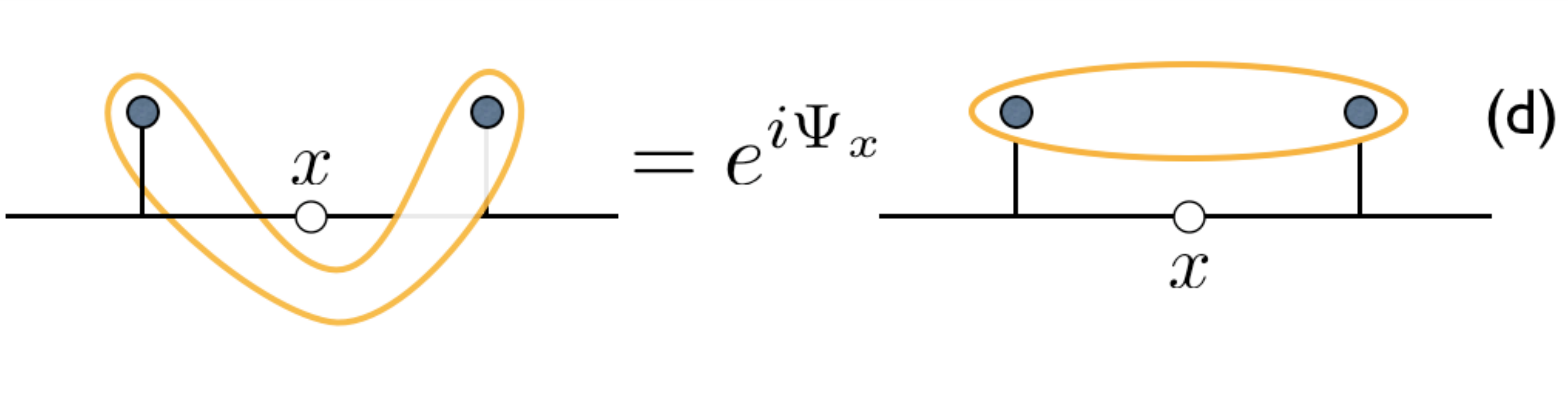}\end{center}
\caption {
(color online)
a) the fusion path connecting $N$ Fibonacci (su(2)$_3$) anyons. Basis states correspond to all admissible labelings of the edges $x_i$ with anyon charges $0$ and $1/2$ that satisfy the fusion rules. b) a nearest neighbor coupling in a chain of anyons can be calculated by using the $F$ matrix  to transform to a new basis in which the fusion product of the two anyons is one of the variables.  c) longer-range interactions first need to be mapped to nearest neighbor interaction by braiding anyons. d) a Dehn twist, giving an additional phase factor is needed if the interaction winds around the torus. For su(2)$_3$ the phase is $\psi_x=0$ for $x=0$ and $\psi_x=-4\pi/5$ for $x=1/2$. 
}
\label{Fig:basis}
\end{figure}

To describe the basis states of a set of $N$ localized (interacting) su(2)$_3$ anyons we consider a fusion path, as shown in Fig. \ref{Fig:basis}(a). The basis of the many-anyon Hilbert space corresponds to all admissible labelings $|x_1,x_2,\ldots\rangle$ of the links in this fusion path with labels $x_i$ corresponding to generalized angular momenta of su(2)$_3$. These labelings must satisfy the constraints of the fusion rules  at each vertex of this fusion path. For su(2)$_3$ these fusion rules are
\begin{eqnarray}
  0   \otimes \alpha  & = & \alpha \nonumber \\
  1/2 \otimes 1/2 & = & 0 \oplus 1 \nonumber \\
  1/2 \otimes 1 & = & 1/2 \oplus 3/2 \nonumber \\
  1/2 \otimes 3/2 & = & 1 \label {Eq:FusionRules} \\
  1 \otimes 1 & = & 0 \oplus 1 \nonumber \\
  1 \otimes 3/2 & = & 1/2 \nonumber \\
  3/2 \otimes 3/2 & = & 0 \nonumber \,,
\end{eqnarray}
where $\alpha \otimes \beta = \beta \otimes \alpha$.
These fusion rules \eqref{Eq:FusionRules} reveal an automorphism $\alpha \rightarrow \hat{\alpha} = 3/2 - \alpha$, allowing an identification of $0 \leftrightarrow 3/2$ and $1/2 \leftrightarrow 1$ for su(2)$_3$. Using the notation for the Fibonacci theory, we write the identity $\bf 1$ for the former and the label $\tau$ for the latter, thus leading to the fusion rules
\begin{eqnarray}
 {\bf 1} \otimes {\bf 1} & = & {\bf 1} \nonumber \\
 {\bf 1} \otimes \tau & = & \tau \label{Eq:FibonacciRules} \\
 \tau \otimes \tau & = & {\bf 1} \oplus \tau \nonumber \,.
\end{eqnarray}
For the labelings of the fusion path these rules then imply that $x_i={\bf 1}$ has to be followed by $x_{i+1}=\tau$ but $x_i=\tau$ can be followed by either  $x_{i+1}={\bf 1}$ or  $x_{i+1}=\tau$.
This constraint gives an overall Hilbert space  size to $F_{N+1}+F_{N-1}\sim \phi^{N}$ (for large $N$) where $F$ is the Fibonacci sequence and $\phi=({1+\sqrt{5}})/{2}$, the golden mean. Note that in comparison to ordinary SU(2) spin-1/2 systems  this Hilbert space has a reduced size.
 
Our specific choice of a fusion path for the $W$-leg ladder system is shown in Fig.~\ref{Fig:Ladder}. Using periodic boundary conditions along the leg direction enables us to conveniently use the translation symmetry of the system along the legs: 
The Hamiltonian matrix can then be block-diagonalized into $L$ blocks labeled by the total momentum $K=2\pi\frac{n}{L}$ of the eigenstates. 
Hence, the Hilbert space (in each symmetry sector) grows approximately as $\phi^N/L$ which is one of the limiting factors of our simulations. To provide some examples, the Hilbert spaces of the $K=0$ sector for $2\times 21$, $3\times 12$ and $4\times 8$ ladders are found to be of sizes $28\, 527\, 448$, $2\, 782\, 659$ and $609\, 147$, respectively.

\subsection {The rung interactions}

\begin{figure}[t]
\begin{center}
	\includegraphics[width=0.9\columnwidth]{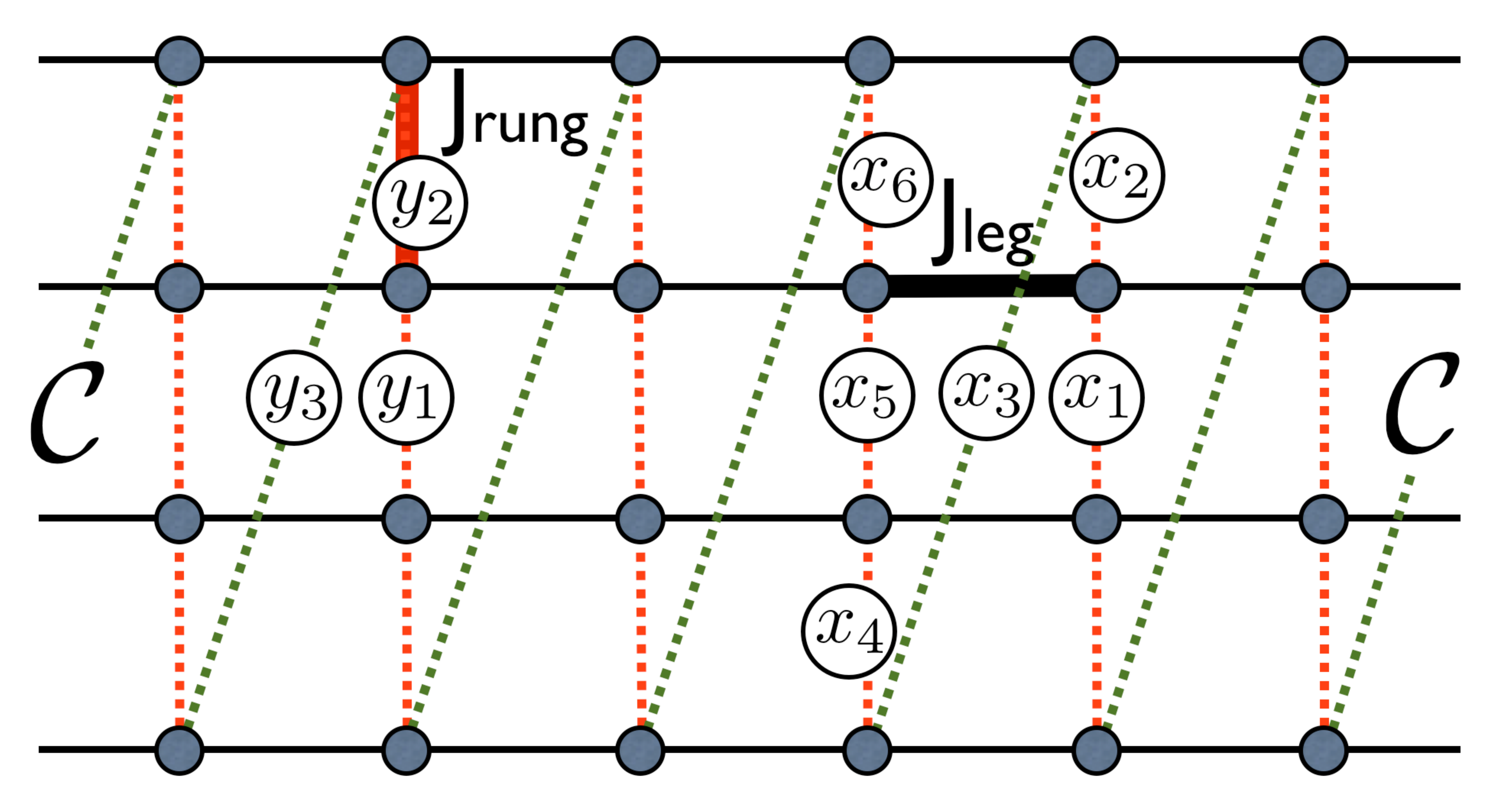}
\end{center}
\caption{
   (color online) 
   Sketch of a typical ladder system of extent $4\times L$ ($L=6$), 
   where the dots indicate the location of $\tau$-anyons.
   The fusion path $\cal C$ (dotted line) with labelings $x_i$ (or $y_i$) is used define
   an (arbitrary) ordering of the sites, which is used in the definition of the basis states.
   The exchange couplings $J_{\rm leg}$ and $J_{\rm rung}$ are indicated.
} 
\label{Fig:Ladder} 
\end{figure}

With our choice of  fusion path the rung coupling $J_{\rm rung}$ on ladders with open boundary conditions on the rungs always connects neighboring anyons along the fusion path.
To calculate the  interaction between two neighboring anyons as in Fig.~\ref{Fig:basis}(b) we need to calculate their total spin by performing a basis transformation using the $F$-matrix, and then assign energy $-J_{\rm rung}$ to the identity fusion channel and energy $0$ to the $\tau$ fusion channel. This basis transformation is illustrated in Figure  \ref{Fig:basis}. 

Denoting the local basis states on the three edges around the interaction as  
$|x_{i-1},x_i,x_{i+1}\rangle\in\{|{\bf 1},\tau,{\bf 1}\rangle, $ $ |{\bf 1},\tau,\tau\rangle,$ $ |\tau,\tau,{\bf 1}\rangle,$ $ |\tau,{\bf 1},\tau\rangle,$ $ |\tau,\tau,\tau\rangle\}$
and the states after the $F$-transformation as
 $|x_{i-1},z_i,x_{i+1}\rangle\in\{|{\bf 1},{\bf 1},{\bf 1}\rangle,$ $|{\bf 1},\tau,\tau\rangle,$ $|\tau,\tau,{\bf 1}\rangle, $ $|\tau,{\bf 1},\tau\rangle, $ $|\tau,\tau,\tau\rangle\}$ we can write the $F$-matrix as
\begin{equation}
 F_i=\begin{bmatrix}
1&0&0&0&0\\
0&1&0&0&0\\
0&0&1&0&0\\
0&0&0&1/\phi&1/\sqrt{\phi}\\
0&0&0&1/\sqrt{\phi}&-1/\phi
\end{bmatrix} \, .
\end{equation}

Assigning an energy $-1$ to the identitiy $z_i={\bf 1}$ and 0 to  $z'_i=\tau$ 
the local Hamiltonian is ${\cal H}_i = -F_iP_iF_i$ where $P_i$ is the projector onto the state with $z_i={\bf 1}$. In the basis defined above we get
\begin{equation}
 {\cal H}_i=\begin{bmatrix}
-1&0&0&0&0\\
0&0&0&0&0\\
0&0&0&0&0\\
0&0&0&-1/\phi^2&-1/\phi^{3/2}\\
0&0&0&-1/\phi^{3/2}&-1/\phi
\end{bmatrix}.
\end{equation}

The rung Hamiltonian is obtained by multiplying this matrix by $J_{\rm rung}$ and, for the term shown in Fig. \ref{Fig:Ladder}  acts on the local states $|y_1,y_2,y_3\rangle$.  


\subsection{The leg interactions}

The leg couplings shown on Fig.~\ref{Fig:Ladder}, on the other hand, are longer range interactions and requires `braiding' of anyons as illustrated in Fig. \ref{Fig:basis}(c). Let us first consider a next-nearest neighbor interaction along a chain. To transform this into a nearest neighbor interaction we need to change the basis again, this time by braiding the two left anyons in a clock-wise manner with a so-called  braid matrix 
$B_i$ acting on the states $|x_{i-1},x_i, x_{i+1}\rangle$. Using the same basis as before this braid matrix can be written as:
\begin{widetext}\begin{equation}
 B_i=\begin{bmatrix}
e^{4i\pi/5}&0&0&0&0\\
0&e^{-3i\pi/5}&0&0&0\\
0&0&e^{-3i\pi/5}&0&0\\
0&0&0&\frac{1}{\phi^2}e^{4i\pi/5}+\frac{1}{\phi}e^{-3i\pi/5}&\frac{1}{\phi^{3/2}}(e^{4i\pi/5}-e^{-3i\pi/5})\\
0&0&0&\frac{1}{\phi^{3/2}}(e^{4i\pi/5}-e^{-3i\pi/5})&\frac{1}{\phi^2}e^{-3i\pi/5}+\frac{1}{\phi}e^{4i\pi/5}
\end{bmatrix}
\end{equation}\end{widetext}
and the next nearest neighbor coupling then becomes $B_i^\dag H_{i+1}B_i$. 

Similarly, for the leg coupling, illustrated for a four-leg ladder in Fig.~\ref{Fig:Ladder}, we need three braids and act on the whole sequence $|x_1, x_2, x_3, x_4, x_5, x_6\rangle$ involving 6 bonds (in general, involving $2W-2$ bonds for W chains) along the fusion path $\cal C$
according to the linear transformation:
\begin{equation}
H_{\rm leg}^{1-6}=  J_{\rm leg} \, B_2 \otimes B_5 \otimes B_4 \otimes {\cal H}_4 \otimes B_4^\dagger \otimes B_5^\dagger \otimes B_2^\dagger  \, .
\label{Eq:Jleg}
\end{equation}
Similar formulas can easily be derived for any bond and any width $W$. 

It should be noticed that acting on any given initial state $|x_1 x_2 \cdots x_{LW}\rangle$,
each of the $W\cdot L$ leg couplings can potentially generate up to $2^{2W-1}$ resulting linear independent states since each operator in \eqref{Eq:Jleg} can generate up to two such states. This exponentially growing number of resulting states should be contrasted to the single state generated by a spin flip operation  in the case of ordinary SU(2) spins. As a consequence, this leads to denser and denser matrices for increasing $W$ in the anyonic ladder models, which limits the numerically accessible system sizes for larger width $W$. 

\subsection{Periodic boundary conditions along the rungs}
\label{periodic}
 
\begin{figure}[t]
\begin{center}
	\includegraphics[width=0.9\columnwidth]{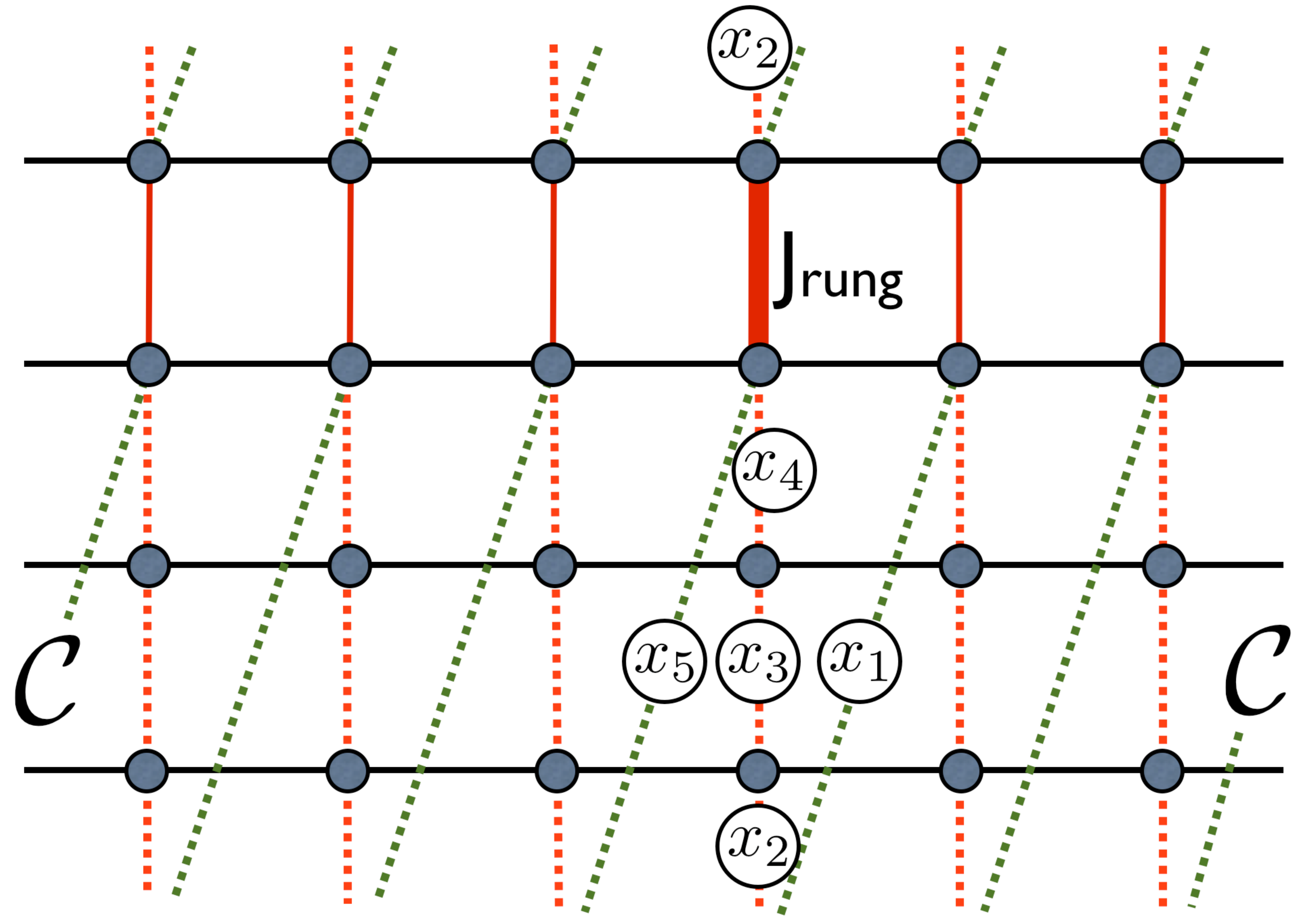}
\end{center}
\caption{
   (color online) 
   A ladder system as in Fig.~\ref{Fig:Ladder}, but with periodic boundary conditions, where an
   an additional rung coupling term connects the two out legs (vertical red segments).
} 
\label{Fig:Ladder2} 
\end{figure}

Closing these open boundaries along the rung direction is done by adding additional couplings between the first and last legs as shown in Fig.~\ref{Fig:Ladder2}. To calculate the Hamiltonian matrix for these couplings one first has to again braid the two involved anyons until they are nearest neighbors along the fusion path. The subtlety with this term is that after the braidings we do not end up with the usual nearest neighbor term of Fig. \ref{Fig:basis}(b) but with a coupling that twists once around the fusion path as illustrated in Fig. \ref{Fig:basis}(d). Untwisting this winding by a $2\pi$ rotation of the right anyon and all following ones by $2\pi$ around the fusion path gives rise to a Dehn twist phase factor $\exp(i\Psi_{x})$~\cite{DehnTwist}, which is $1$ for $x={\bf 1}$ but $\exp(-4i\pi/5)$ for $x=\tau$.

The Hamitlonian for this rung term acts on the local sites $x_1, x_2, x_3, x_4, x_5$ of Fig. \ref{Fig:Ladder2} and reads 
\begin{equation}
{\tilde H}_{\rm rung}^{1-5}=  J_{\rm rung} \, B_2 \otimes B_3 \otimes T_4 \otimes {\cal H}_4 \otimes T_4^\dagger \otimes B_3^\dagger \otimes B_2^\dagger  \, ,
\label{Eq:Jrung2}
\end{equation}
where $T_i$ is the (diagonal) $2\times 2$ twist matrix, 
\begin{equation}
 T_i=\begin{bmatrix}
1&0\\
0&e^{-4i\pi/5}\\
\end{bmatrix}\, ,
\end{equation}
in the local $\{|{\bf 1}\rangle,|\tau\rangle\}$ basis (for the variable $x_i$). 

Care must be taken in choosing a consistent convention for the phase of  (counter)-clock wise braids and Dehn twists. An inconsistent choice can easily be detected as it will cause a broken translation symmetry along the rungs that can be seen in, {\it e.g.} the local bond energies. 

\section{Strong coupling limit and phase diagrams}
\label{Sec:StrongCoupling}

\begin{figure}
\begin{center}
	\includegraphics[width=\columnwidth]{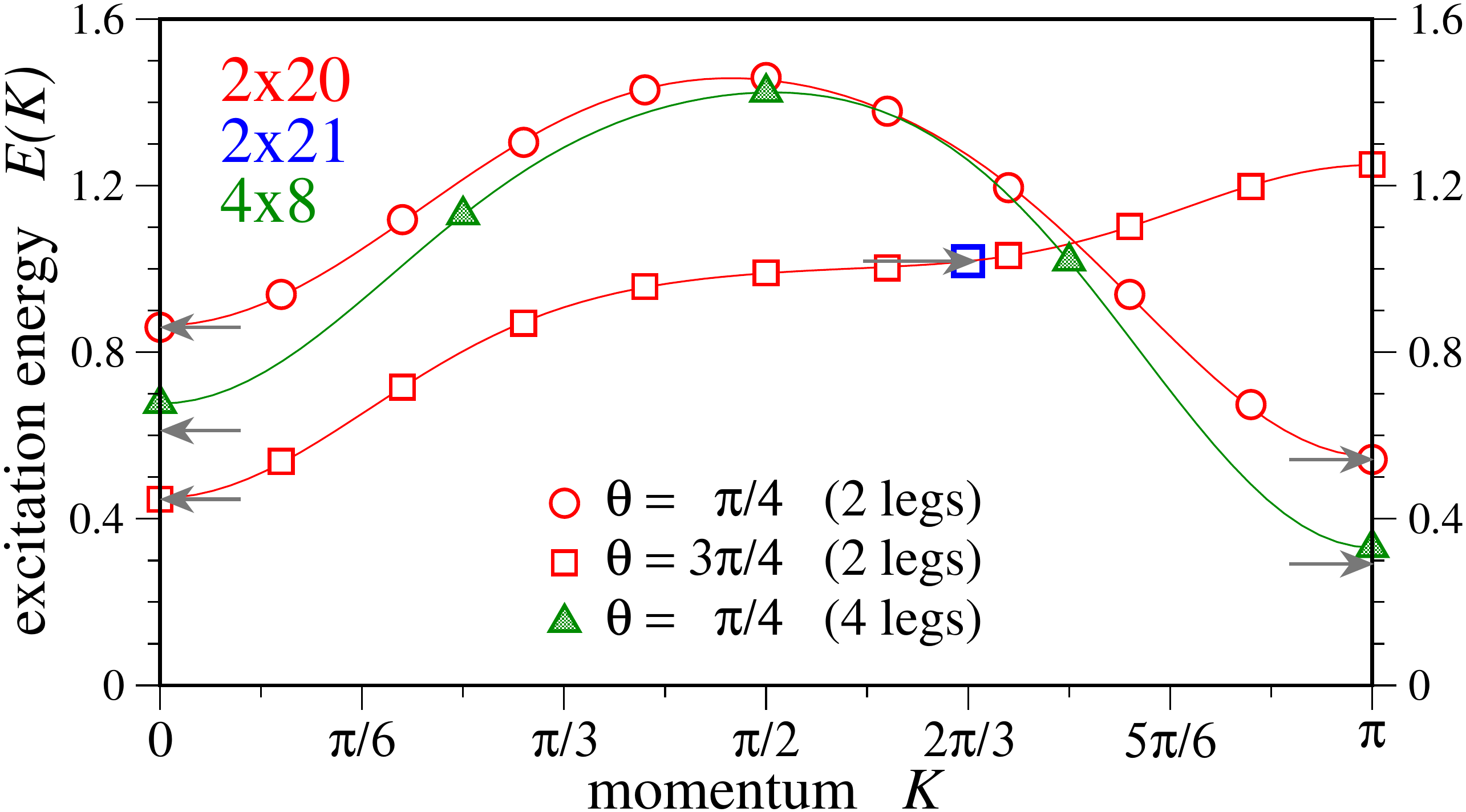}
\end{center}
\caption {(color online) 
	Dispersion of a generalized `magnon' excitation along the ladder direction for two-leg and
	four-leg ladders and various couplings. 
	The magnon excitation is created by flipping a local label $x_i$ from $\tau$ to $\bf 1$,
	which can then propagate down the ladder.
	Data is shown for $2\times 20$ (and $2\times 21$) ladders (open symbols)
	as well as $4\times 8$ ladder (closed symbols) and coupling parameters
	$\theta=\pi/4$ ($J_{\rm leg}=J_{\rm rung}=\sqrt{1/2}$) and 
	$\theta=3\pi/4$ ($J_{\rm leg}=-J_{\rm rung}=-\sqrt{1/2}$). 
	The solid lines are a guide-to-the-eye, obtained from Fourier series fit to the data.
	The arrows indicate  extrapolations to the thermodynamic limit as shown in 
	Fig.~\protect\ref{fig:gaps}.}
\label{fig:magnon}
\end{figure}

For SU(2) quantum spin ladders a single phase extends from the weak to strong rung coupling limit for any of the four possible signs of the rung and leg couplings \cite{OrdinaryLadders,Greven96}. 
The generic phase diagram thus has at most four different phases. 
For su(2)$_3$ anyonic ladder we observe the same behavior and we will start to discuss the various
phases starting from the strong rung coupling limit $|J_{\rm rung}| \gg |J_{\rm leg}|$. 
The results discussed below are summarized in Table~\ref{table:phase_diagram}. The total spin of an isolated rung, which depends on the sign of the rung coupling and on the 
rung length $W$, completely determines the nature of the phase at finite $J_{\rm leg}$ and whether it is gapped or critical. 
For antiferromagnetic $J_{\rm rung}$  (first two lines of Table~\ref{table:phase_diagram}) we find similar 
even/odd effects as in the SU(2) case. Even widths are gapped while odd widths are critical and characterized by the same CFT as the single chain. 
For ferromagnetic  $J_{\rm rung}$ and $W=3p$ ($p$ an integer) the rungs 
form singlets, (labeled with the identity $\bf 1$) and hence, the ladders are gapped.
Otherwise, the rungs behave as ``triplet'' ($\tau$) states and the low-energy physics is that of an (effective) critical chain as shown in the two last lines of Table~\ref{table:phase_diagram}.

\begin{table}
\begin{center}
 \begin{tabular}{@{} ccccccc @{}}
   \toprule
   W$\rightarrow$& 1 & 2 & 3 & 4 & 5 & 6\\ 
   \hline
   AFM-AFM& 7/10 & $\Delta$ & 7/10 & $\Delta$&7/10& $\Delta$\\ 
   AFM-FM& 4/5 & $\Delta$& 4/5 & $\Delta$& 4/5 & $\Delta$ \\ 
   FM-AFM & 7/10 & 7/10 & $\Delta$& 7/10 & 7/10 & $\Delta$ \\ 
   FM-FM & 4/5 & 4/5 & $\Delta$ & 4/5& 4/5 & $\Delta$ \\ 
   \toprule
 \end{tabular}
\end{center}
\caption{The various phases of the Fibonacci ladders as a function of the number of legs W. Each line corresponds to one of the four quadrants of the parameter space shown in Fig.\protect\ref{Fig:Circle}. In the first column, 
the first (second) label refers to the rung (leg) coupling $J_{\rm rung}$  ($J_{\rm leg}$). Gapped phases are labeled by $\Delta$. For the gapless phases 
the value of the central charge of the low-energy CFT is indicated.}
\label{table:phase_diagram}
\end{table}

\subsection{Antiferromagnetic rung coupling} 
Let us first consider AFM rung coupling, where for an isolated rung the ground state has total angular momentum $j=0$ (state with label $\tau$) for even width $W$ and $j=\frac{1}{2}$ 
(state with label $\bf 1$) for odd width. 

\begin{figure}
\begin{center}
	\includegraphics[width=\columnwidth]{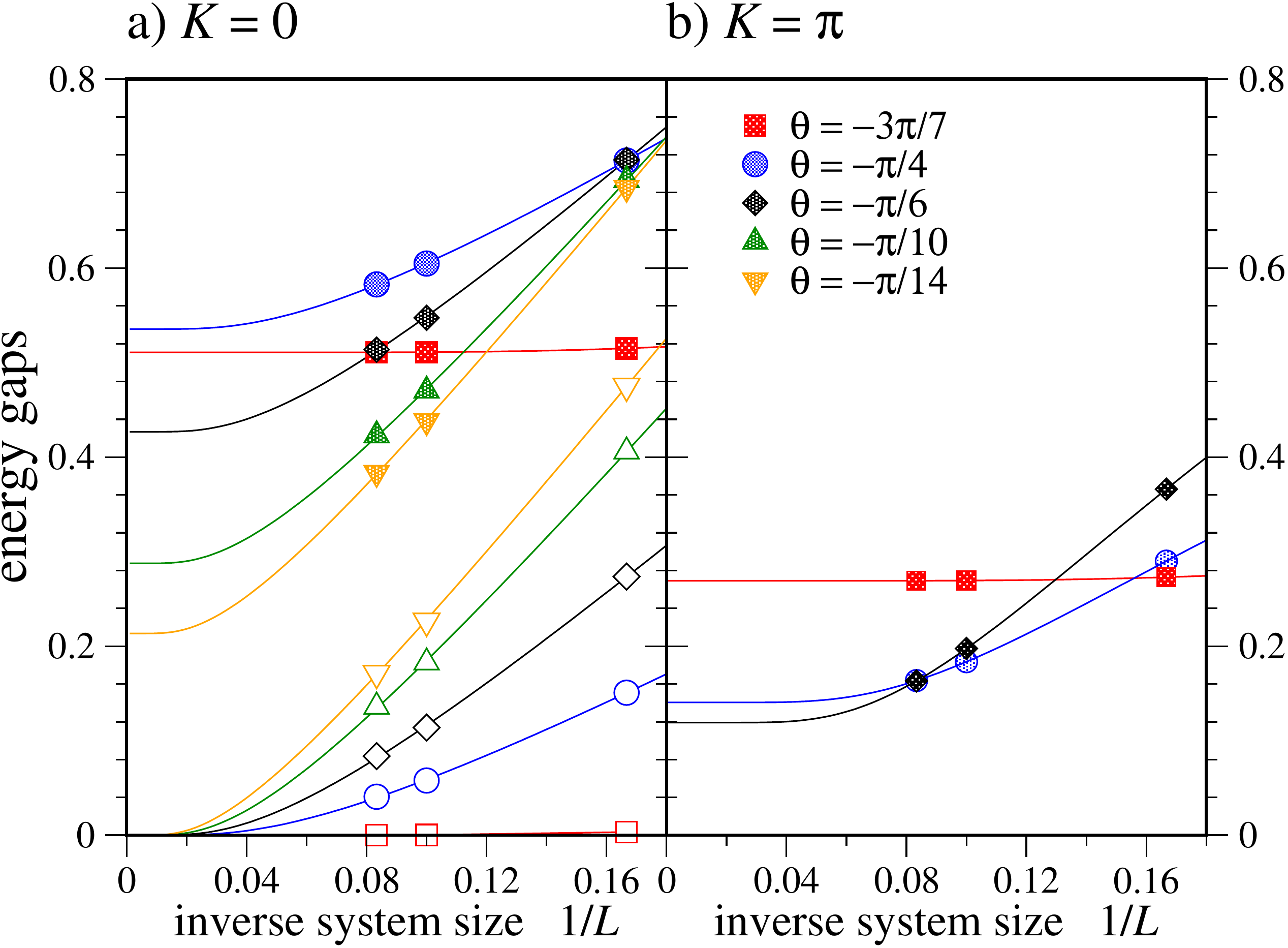}
\end{center}
\caption{
   (color online) 
   Finite-size extrapolations of the energy gaps of 3-leg ladder systems for various couplings $\theta$,
   with ferromagnetic rung and antiferromagnetic leg couplings in the a) $K=0$ and b) $K=\pi$  
   momentum sectors.
   Extrapolations to the thermodynamic limit are obtained by fitting the numerical data to the form
    $\Delta(L)\simeq \Delta(\infty) +\frac{C}{L} \exp{(-L/\xi)}$, where $\xi$ is a correlation length. 
    In the $K=0$ sector, the first excitation energy (open symbols) extrapolates to zero 
    (indicating that the ground state is two-fold degenerate in the thermodynamic limit),
    while the  extrapolation of the second excitation energy (filled symbols) indicated a finite gap.
 }
\label{Fig:Scaling_3legs}
\end{figure}

For even $W$, the ground state at $J_{\rm leg}=0$ is a product of local $\tau$ states on the individual rungs. 
The elementary excitation is a local singlet ($\bf 1$) excitation with a gap   
$\Delta_0(W)\sim 1/W$.
For (weak) leg coupling the elementary `magnon' excitation can hop to one of its two neighboring rungs in first-order in $J_{\rm leg}$, giving rise to a dispersion of width $\propto |J_{\rm leg}|$. 
Typical such dispersions (but for intermediate couplings) are shown in Fig.~\ref{fig:magnon}. 
The gap decreases linearly as  $\Delta_0(W)-\alpha |J_{\rm leg}| +O( J_{\rm leg}^2 )$.
However, this perturbative strong-coupling result for the gap is restricted to a shrinking region $\sim 1/W$ around $J_{\rm leg}=0$ as $W$ gets larger. 
For ordinary SU(2) ladders it has been argued that the gap vanishes as $\exp{(-cW)}$ for large enough $W$ and any chain coupling $J_{\rm leg}$~ \cite{Greven96}.
We will return to the question whether  a gap can survive for anyonic systems in the limit 
$W\rightarrow \infty$ below.

\begin{figure}
\begin{center}
	\includegraphics[width=\columnwidth]{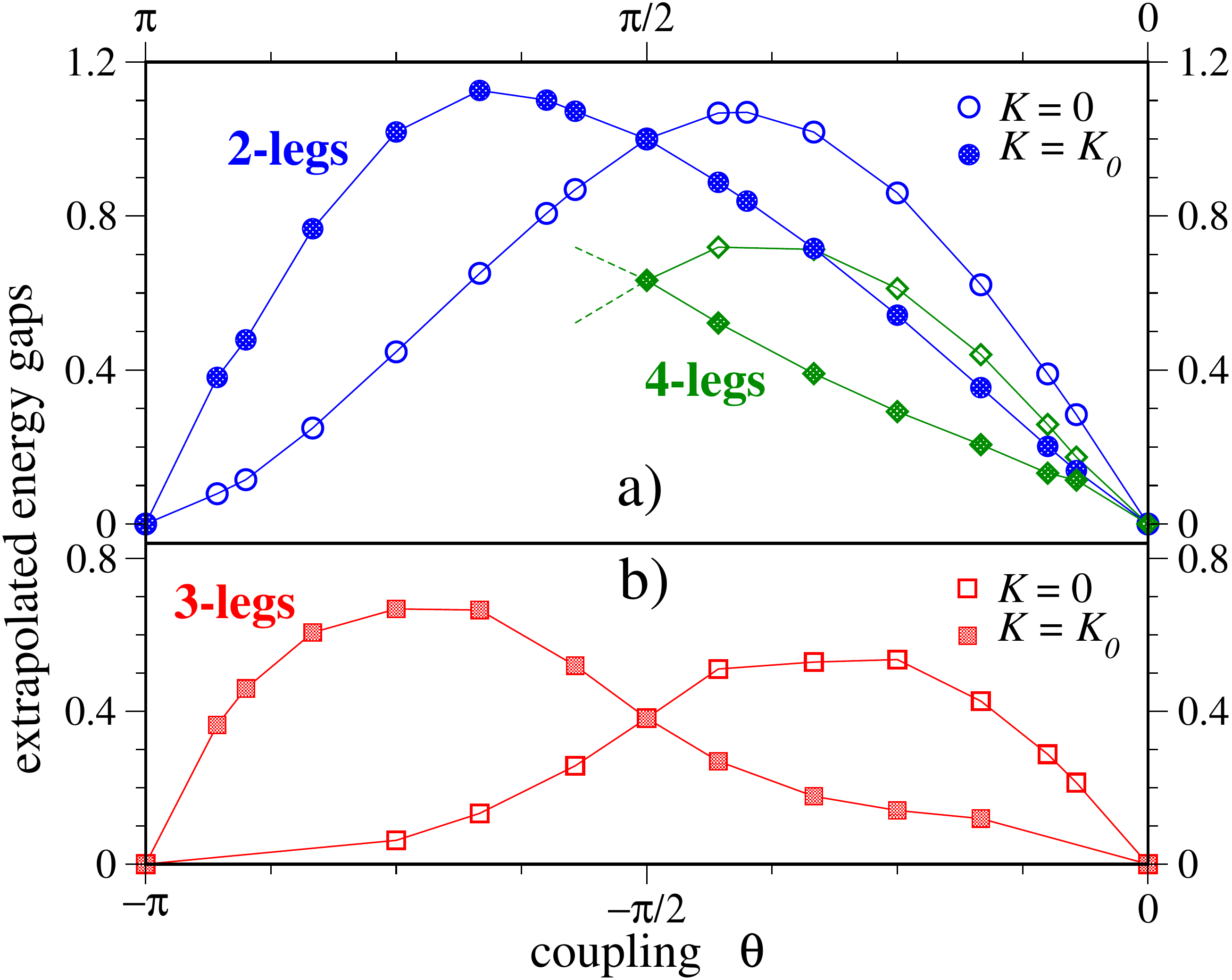}
\end{center}
\caption {(color online)
Extrapolated values $\Delta(L \to \infty)$ of the gaps at the two crystal momenta $K=0$ and $K=K_0$ corresponding to the zero-energy modes of the single chains.
$\Delta(\infty)$ is plotted as a function of coupling parameter $\theta$, 
i.e. $\theta\in [0,\pi ]$ for 2- and 4-leg ladders with AFM rung coupling (a) and
$\theta\in [-\pi,0]$ for a 3-leg ladder with FM rung coupling (b).
For AFM (FM) leg coupling, $K_0=\pi$ ($K_0=2\pi/3$) and clusters up to
$2\times 20$ ($2\times 21$), $3\times 12$ ($3\times 12$) and
$4\times 8$ have been used. $\Delta(\infty)$ is deduced
by  fitting the data as $\Delta(L)\simeq \Delta(\infty) +\frac{C}{L}
\exp{(-L/\xi)}$, where $\xi$ is a correlation length.
} 
\label{fig:gaps}
\end{figure}

Away from the above-discussed limit, the gaps of the 2- and 4-leg ladders can be obtained for intermediate couplings by using Lanczos exact diagonalisations of clusters of different lengths.
Finite size scalings (similar to the one shown in Fig.~\ref{Fig:Scaling_3legs} for a 3-leg ladder to be discussed later) enable to accurately estimate, in the thermodynamic limit,
the gaps at the minima of the dispersion (see e.g. Fig.~\ref{fig:magnon}) of the excitation spectrum. 
Results of the extrapolated gaps are summarized in Fig.~\ref{fig:gaps}(a). Note that the minima of the dispersion occurs at different momenta depending on the sign of $J_{\rm leg}$,
0 and $\pi$ for antiferromagnetic $J_{\rm leg}>0$, 0 and $2\pi/3$ for ferromagnetic $J_{\rm leg}<0$. In the latter case, for sufficiently large leg coupling, the minima at $2\pi/3$ 
can disappear as shown in Fig.~\ref{fig:magnon}.  

\begin{figure}
\begin{center}
	\includegraphics[width=\columnwidth]{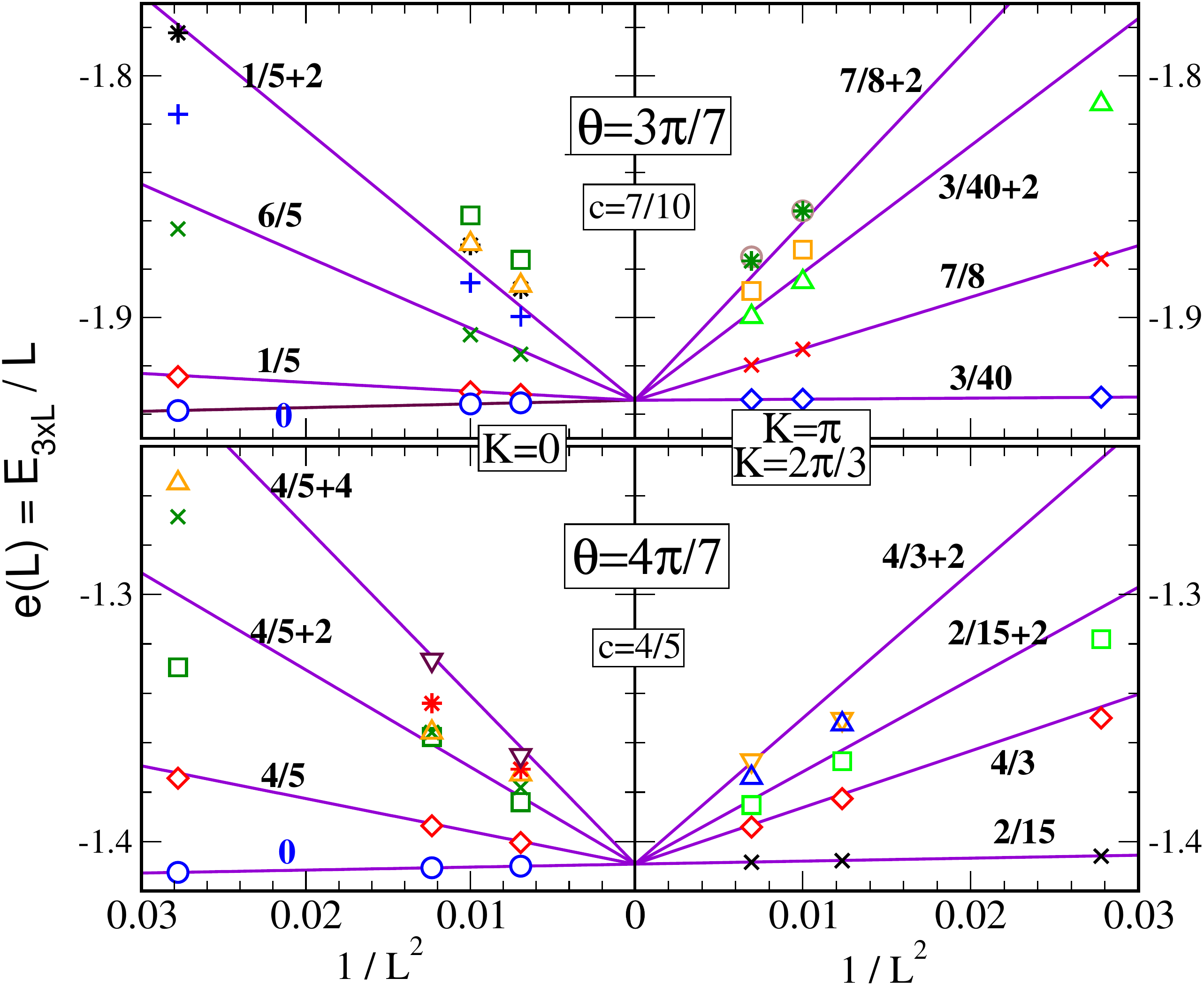}
\end{center}
\caption {(color online)
Finite size scaling 
of $3\times L$   (3-leg) ladders of sizes up to $L=12$, with AFM rung coupling and with both AFM (top) and 
FM (bottom) leg couplings.  The lowest eigen-energies per rung (symbols) are plotted vs $1/L^2$.
The right and left panels correspond to two different crystal momenta, the ones 
characterizing the zero-energy modes of a single chain.
The $\theta$ values (indicated on the plot) correspond to the strong rung couplings $|J_{\rm rung}/J_{\rm leg}|\sim 4.38$ labelled as `1' and `2' in Fig.~\ref{Fig:Circle}.
CFT scalings are shown: (i) a linear fit of the GS energy (labeled by `0') accurately provides the overall energy 
scale (i.e. the velocity); (ii) all other straight lines are expected CFT scalings using the conformal weights $2h_n$ and the central charge $c$ indicated on the plots
(see Eq.~(\ref{CFT_scaling}) in the text). 
} 
\label{fig:str-cpl}
\end{figure}

For odd width $W$, since the GS of a single AFM rung carries angular momentum $j=1/2$ ($\tau$), 
the low-energy effective model for weakly coupled rungs is that of a single $\tau$-anyon chain.
Indeed, as shown in Fig.~\ref{fig:str-cpl} for a 3-leg ladder
we find that the low-energy spectrum is gapless and can be described  by a Conformal Field
Theory  (CFT) identical to the one of a single chain~\cite{Feiguin07}.
In particular, we find that the lowest energies (per rung) $e_n$ scale as
\begin{equation}
e_n(L)\simeq e_\infty + \pi u(-\frac{c}{12}+2h_n)\frac{1}{L^2}\, ,
\label{CFT_scaling}
\end{equation}
where $L$ is the length of the ladder, $u$ is a zero-mode velocity, and $2h_n$ and $c$ are the conformal weights (or scaling dimensions) and central charge of the CFT. Depending on the sign of the leg coupling 
these gapless theories are those of the tricritical Ising model ($c=7/10$) or 3-state Potts model
($c=4/5$) for AFM and FM couplings respectively.

The anyonic ladders with AFM rung coupling thus behave similarly to their SU(2) analogs with  even/odd widths giving rise to gapped/gapless physics as summarized in the
first two lines of Table~\ref{table:phase_diagram}.

\subsection{Ferromagnetic rung coupling} 
Next, we move to the case of a ferromagnetic rung coupling, where we find major differences between the anyonic ladders and their SU(2) counterparts. 
For ordinary SU(2) ladders of even width $W$, the strong coupling rungs form a total integer spin and  the effective low-energy model (for weak leg coupling) is a Haldane Heisenberg chain~\cite{Haldane}
which is gapped for AFM leg coupling. 
For odd width $W$ the ladders remain gapless for either sign of the leg coupling, since each rung forms a $\tau$ state. 

\begin{figure}
\begin{center}
	\includegraphics[width=\columnwidth]{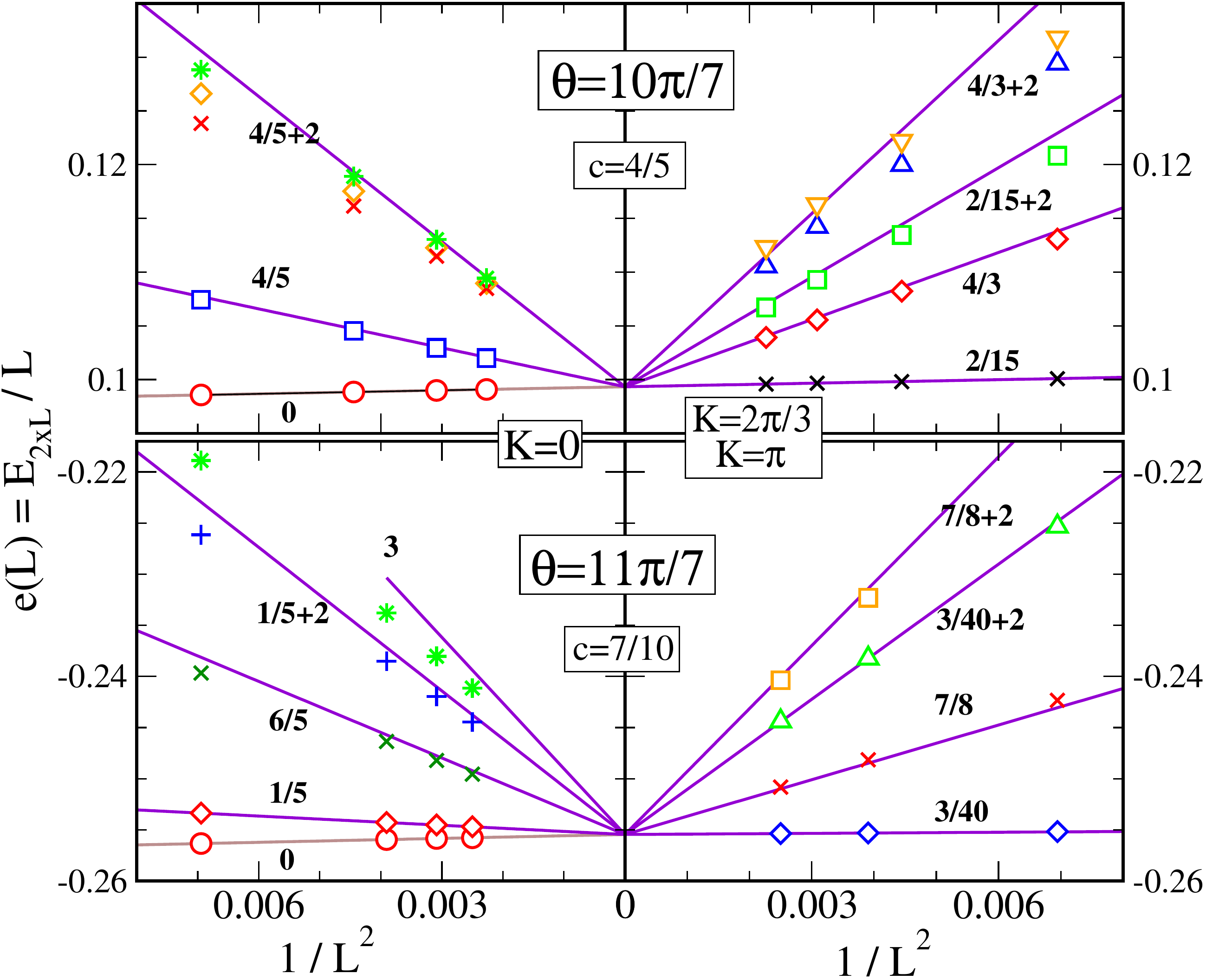}
\end{center}
\caption {(color online)
Finite size scaling  of  $2\times L$  (2-leg) ladders with FM rung coupling, with both FM (top) and AFM (bottom)
leg couplings and with up to $2\times 21$ and $2\times 20$ sites, respectively.  
The $\theta$ values (indicated on the plot) correspond to the strong rung couplings $|J_{\rm rung}/J_{\rm leg}|\sim 4.38$ labelled as `3' and `4' in Fig.~\ref{Fig:Circle}.
Same notations and same analysis as in Fig.~\ref{fig:str-cpl}.} 
\label{fig:str-cpl2}
\end{figure}

Also in contrast to ordinary SU(2) spins, for anyonic ladders, we find different phases and periodicities of 3 in $W$ for $k=3$.
Ladders with $W=3p$ ($p$ an integer) and ferromagnetic $J_{\rm rung}$ are gapped, since each rung forms a singlet ($\bf 1$) state similar to the even width ladders in the AFM case.  As an example, we show in Fig.~\ref{fig:gaps}(b) the gap of a 3-leg ladder obtained from finite size scalings, examples of which are shown in Fig.~\ref{Fig:Scaling_3legs}.
Alternatively, the low-energy effective model of ladders with widths that are not multiples of 3 is again that of a single $\tau$-chain and thus gapless as illustrated in Fig.~\ref{fig:str-cpl2} for a 2-leg ladder. 
One might naively expect that the 2-leg ladder is again a gapped Haldane chain, since two FM coupled $j=1/2$ momenta form a total $j=1$ momentum. However, as noted in the introduction, in su(2)$_k$ theories with odd level $k$ one can identify momentum $j$ with momentum $k/2-j$ by fusing it with the Abelian momentum-$k/2$ particle. For su(2)$_3$ this implies that momentum $j=1$ behaves like momentum 
$j=1/2$ ($\tau$). We find that this gapless phase extends all the way up to weak rung coupling.

\begin{figure}
\begin{center}
	\includegraphics[width=\columnwidth,angle=0]{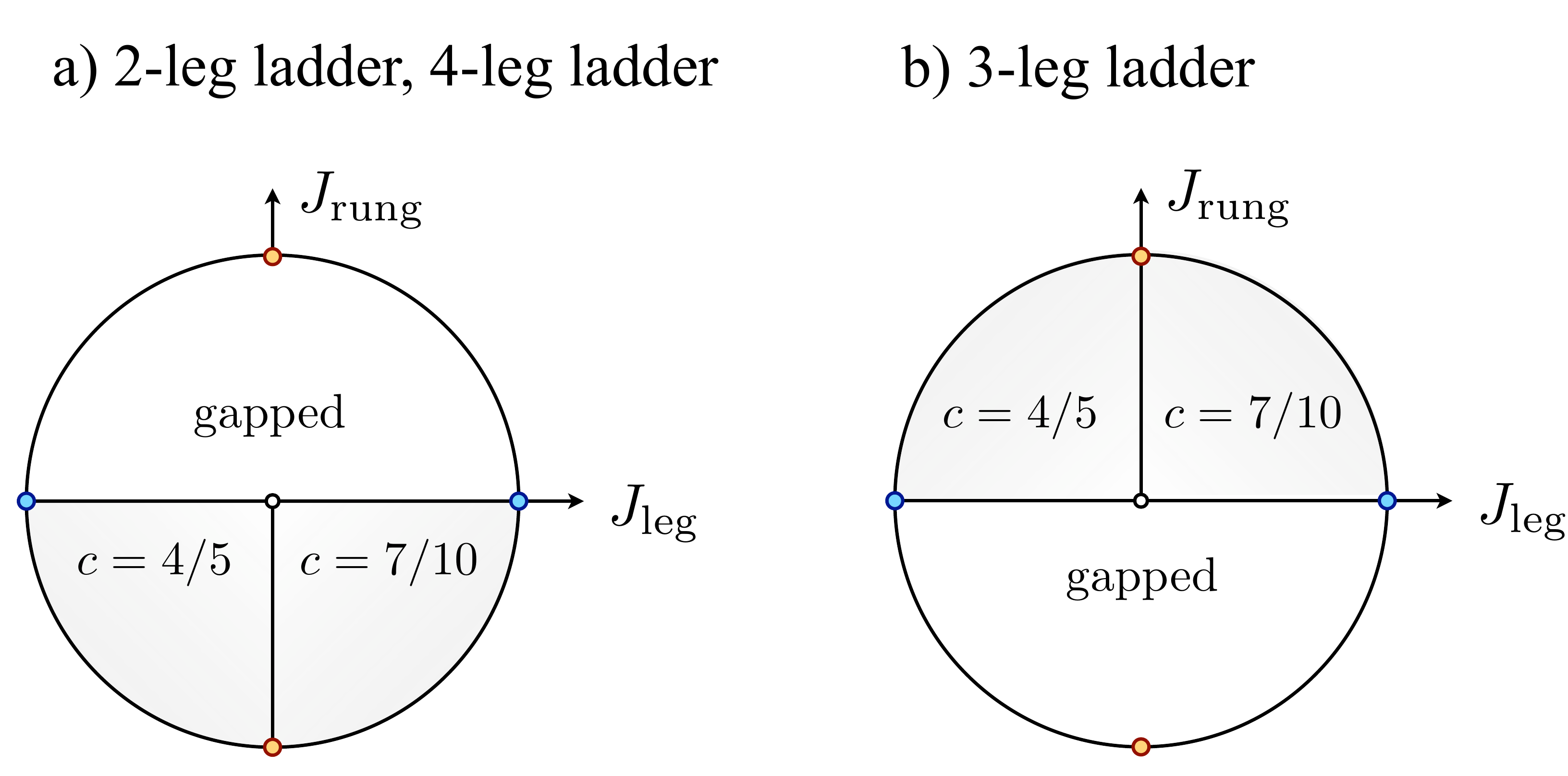}
\end{center}
\caption {(color online)
Phase diagrams of the 2-leg and 4-leg ladders (a) and of the 3-leg ladders (b) versus the couplings
$J_{\rm leg}=\cos{\theta}$ and $J_{\rm rung}=\sin{\theta}$. The central charge of the gapless phases is
indicated. 
} 
\label{Fig:Phasediags}
\end{figure}

We summarize our results in the phase diagrams of Fig.~\ref{Fig:Phasediags}(a) for 2-leg and 4-leg ladders, and of Fig.~\ref{Fig:Phasediags}(b) for 3-leg ladders. 

\section{Decoupled chains} 
\label{Sec:Liquids}

\begin{figure}
\begin{center}
	\includegraphics[width=\columnwidth]{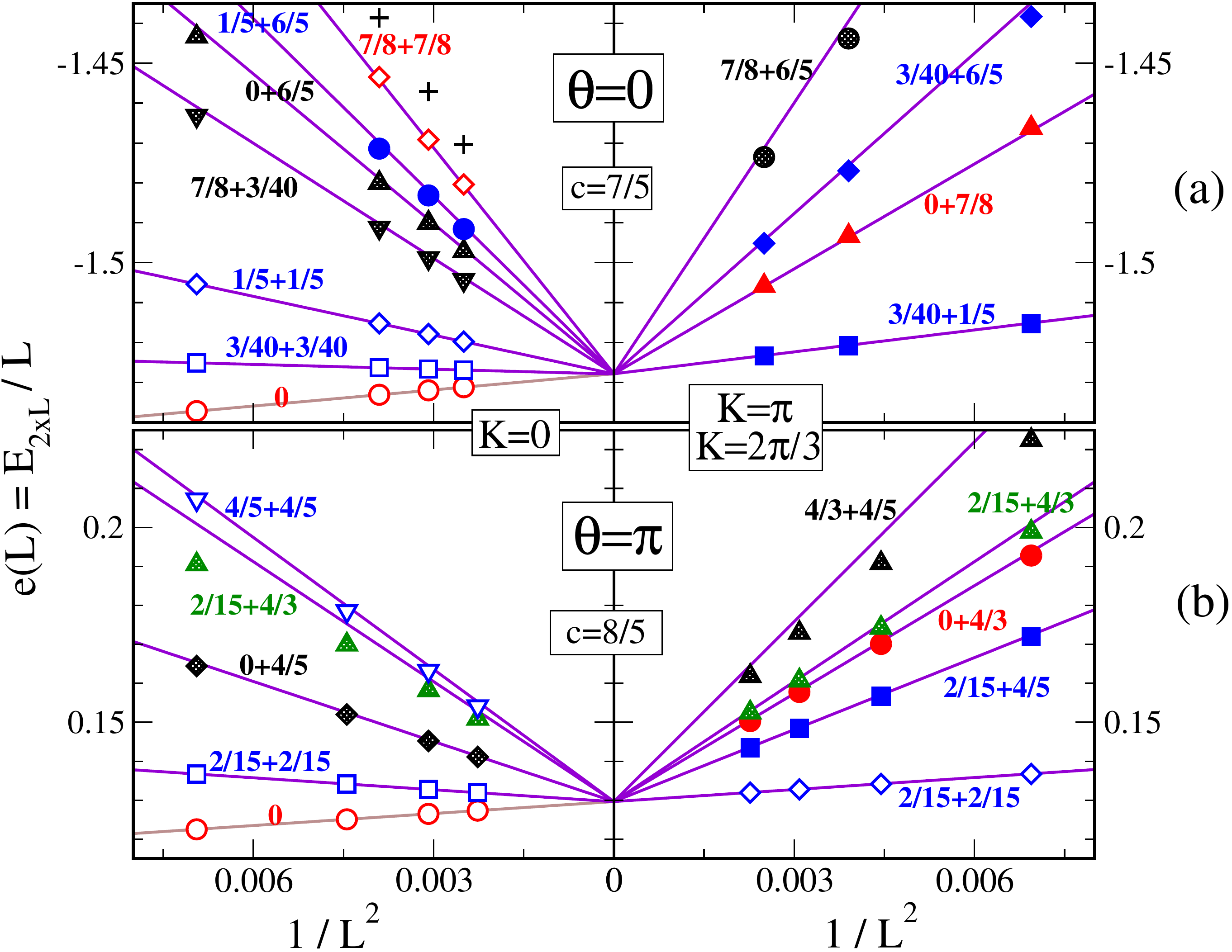}
\end{center}
\caption {(color online)
Finite size scaling of the lowest eigen-energies (normalized per rung)  of a 2-leg ladder with $J_{\rm rung}=0$ and $J_{\rm leg}=\pm 1$ (decoupled chains) vs $1/L^2$.
The right and left panels correspond to two different crystal momenta, the ones 
characterizing the zero-energy modes of a single chain.
Same procedure and notations as Fig.~\ref{fig:str-cpl} and Fig.~\ref{fig:str-cpl2}. 
} 
\label{fig:axis}
\end{figure}

We now turn to a discussion of the limit where the rung coupling between the
individual legs of the ladder vanishes. 
In contrast to  the case of conventional SU(2) spin ladders, we find that the 
anyonic ladder system does {\sl not} decompose into independent chains
in  this limit of vanishing rung coupling, i.e. $J_{\rm rung}=0$. In particular, 
we find that the energy spectrum in this limit is {\sl not} given by the free tensor 
product of the energy spectra of individual chains, but rather turns out to be 
a certain subset thereof.
In the following, we describe a set of `topological gluing conditions' that 
constrain the energy spectrum to this subset of the free tensor product. We
closely follow the analytical arguments, which we developed in 
Refs.~\onlinecite{Gils09,Ludwig10} in a so-called `liquids picture', where
we identify the collective gapless modes of the quasi one-dimensional anyon 
chains (or ladders) with edge states at the spatial interface between two distinct topological 
quantum liquids -- for an illustration see e.g. Fig.~2 of Ref.~\onlinecite{Gils09}.
This `liquids picture' provides a set of analytical rules which allow to obtain the 
spectrum of these decoupled anyon chains, which in the remainder of this section
we compare with numerical results for 2-leg and 3-leg ladder systems.
We find perfect agreement of the two approaches.


\begin{table}[t]
\begin{center}
 {\bf 2-leg ladder\\}
 \begin{tabular}{@{} |c|c||c|c||c|c||c|c| @{}}
   \toprule
   $\theta = 0$ &  top. &$\theta = 0$  & top. &$\theta = \pi$& top.  &$\theta = \pi$& top. \\
 $K = 0$ & sector & $K = \pi$  & sector  & $K = 0$ & sector  & $K= \pi/3$ &sector\\ 
   \hline 
   I & 0 & $\sigma+\epsilon$ & ($\frac{1}{2}$,$\frac{1}{2}$) & $I$ & 0 & $\sigma+\sigma$ & $\frac{1}{2}$ \\

   $\sigma + \sigma$ & $\frac{1}{2}$ & $I+\sigma'$ & 0 & $\sigma+\sigma$ & $\frac{1}{2}$ & $\sigma+\epsilon$ &($\frac{1}{2}$,$\frac{1}{2}$) \\ 

    $\epsilon + \epsilon$ & $\frac{1}{2}$ & $\sigma + \epsilon'$ & ($\frac{1}{2}$,$\frac{1}{2}$) & $I+\epsilon$ & (0,$\frac{1}{2}$) & $I+\psi$& (0,0) \\  

   $\sigma + \sigma'$ & (0,$\frac{1}{2}$) & $\sigma' + \epsilon'$ & (0,$\frac{1}{2}$) & $\sigma+\psi$ & (0,0,$\frac{1}{2}$,$\frac{1}{2}$)  & $\sigma+\psi$ & (0,0,$\frac{1}{2}$,$\frac{1}{2}$) \\ 

   $I+\epsilon'$ & (0,$\frac{1}{2}$) & ´ & ´ & $\epsilon+\epsilon$ & $\frac{1}{2}$ & $\epsilon+\psi$ & (0,$\frac{1}{2}$) \\ 

$\epsilon + \epsilon'$ & ($\frac{1}{2}$,$\frac{1}{2}$) & ´ & ´ & ´ & ´ & ´ & ´ \\ 

$\sigma' + \sigma'$ & 0 & ´ & ´ & ´ & ´ & ´ & ´ \\ 
   \toprule
 \end{tabular}
\end{center}

\begin{center}
 {\bf 3-leg ladder\\}
 \begin{tabular}{@{} |cl||c||lc||lcl| @{}}
   \toprule
   $\theta=0$ & ´&  $\theta = 0$ & ´ &$\theta = \pi$ & ´ &$\theta = \pi$ & ´\\
 $K=0$ & ´& $K=\pi$ & ´ & $K=0$& ´&  $K=2\pi/3$ & ´ \\ 
   \hline 
   $I$ & ´ &   $\sigma+\sigma+\sigma$ & ´ &  0 & ´ &  $\sigma+\sigma+\sigma$ & ´ \\ 

 $\sigma+\sigma+\epsilon$   & ´ &  $\epsilon + \epsilon +\sigma$ & ´ &  $I+\sigma+\sigma$ & ´ &  $I+\sigma+\epsilon$ & ´ \\ 

    $\epsilon+\epsilon+\epsilon$ & ´ &  $I+I+\sigma'$& ´ &  $I+I+\epsilon$ & ´ & $I+I+\psi$ & ´ \\ 

   $I+\sigma+\sigma'$ & ´ & ´ & ´ & $\sigma+\sigma+\epsilon$ & ´ & ´ & ´ \\ 
   \toprule
 \end{tabular}
\end{center}
\caption{
	Scaling dimensions of for two (top) and three (bottom) decoupled chains with 
	antiferromagnetic  ($\theta=0$) and ferromagnetic ($\theta=\pi$) leg couplings, 
	for the two crystal momenta corresponding to the respective zero-energy modes of a single chain. 
	Listed here are the numerically observed conformal operators in the subset of the free tensor 
	product of energy states for $W$ individual chains, corresponding to conformal field theories with
	central charge $c=W\times\frac{7}{10}$ ($c=W\times\frac{4}{5}$) for antiferromagnetic
        (ferromagnetic) coupling.
        For the Ising theory ($c=7/10$), we use the common identification of operators with conformal 
        weights, 
        i.e. $I \to 0, \epsilon \to 1/5, \epsilon' \to 6/5, \epsilon'' \to 3, \sigma \to 3/40, \sigma' \to 7/8$.
        For the 3-state Potts model ($c=4/5$) this identification becomes
        $I \to 0, \epsilon \to 4/5, \sigma_{1,2} \to 2/15, \psi_{1,2} \to 4/3$.
	For the case of two chains we also list the topological symmetry sector, i.e. the eigenvalue
	of the topological symmetry operator, for all energies.}
\label{table:decoupled_legs}
\end{table}

Let us briefly describe the analytical spectrum of the
decoupled chains
($J_{\rm rung}=0$ limit), based on the results obtained in 
Ref.~\onlinecite{Gils09} (see Fig. 2 of that reference).
At each interface between two topological (or Hall) liquids
there is an edge (see Fig.~\ref{Fig:Circle}(b) of Ref.~\onlinecite{Gils09}). 
The key tool developed in Ref.~\onlinecite{Gils09} was that 
each chain can be viewed as `filled' with a new topological 
(or Hall) liquid so that the right- and left- moving gapless degrees of freedom
of each chain arise from the juxtaposition of two topological liquids.
The field theory describing each of these edges
arises from the familiar
GKO coset construction \cite{GKO-REF} of conformal field theory.
Consider for example two decoupled chains. Thus, there
are five liquids, and four edges.
For, say,  "antiferromagnetic (AF)" interactions between the anyons, the spectrum 
of these four edges in the `topological sector' of `topological charge' $j_1$
takes on the form 
(following the rules developed in Ref.~\onlinecite{Gils09},
and using the notation of the same article) 
$$
(\psi_L)^{j_1}_{j_2} \ (\psi_R)_{j_2}^{j_1'} \ (\psi_L)^{j_1'}_{j_2'} 
\ (\psi_R)_{j_2'}^{j_1}
$$
Here $j_1$ denotes the `topological charge' which is  "ejected" from
the four-edge system to infinity through the surrounding ("parent")
topological liquid
[compare again Fig.~\ref{Fig:Circle}(b) of Ref.~\onlinecite{Gils09}]. 
The left (holomorphic) and right (antiholomorphic) conformal weights
of this state are
\begin{equation}
\label{hLeft}
h_L= \Delta^{j_1}_{j_2} \ + 0+ \Delta^{j_1'}_{j_2'} +0
\end{equation}
\begin{equation}
\label{hRight}
h_R=  0 + \Delta_{j_2}^{j_1'} \ +0  +\Delta_{j_2'}^{j_1}
\end{equation}
respectively,
where
$
\Delta^{j_1}_{j_2}
=
[1 + 
j_2 (j_2 +1)/(k+1) -
j_1 (j_1 +1)/(k+2)]
$
is the conformal weight of the primary field
$(\psi_L)^{j_1}_{j_2}$
in the GKO coset
${\rm su(2)}_{k-1}\times {\rm su(2)}_1/{\rm su(2)}_k$.
Considering for simplicity $k=$ odd, 
we can choose $j_1$ and $j_1'$ to run over integer
values $0, 1, ..., (k-1)/2$ and $j_2, j_2'$ run over values
$0,1/2, 1,..,(k-1)/2$.
We are only interested in fields with $h_L=h_R$, so
that the scaling dimension is $x(j_1;j_2;j_1';j_2';j_1) = 2 h_L$.

We now compare the above analytical spectrum with numerical
spectra for decoupled chains.
Finite size scaling  of the low-energy spectrum
(similar to the procedure employed in previous Section) 
enables us to 
assign conformal weights to each energy level,
analogous to the case of the 2-leg su(2)$_3$ ladder 
in Fig.~\ref{fig:axis}. As before, an accurate fit of the
groundstate (GS) energy per site versus $1/L^2$ fixes the overall energy scale.
The (allowed) combinations $2h_L=2h_1+2h_2$ 
corresponding to the sum of
two  conformal weights,
each arising from a single edge state
(compare Eq. \ref{hLeft}),
can then be read off from the slopes of the 
lowest excited states versus $1/L^2$. These numerical results as well as 
those for three decoupled legs
(scaling not shown here), and for both ferro 
and antiferromagnetic (intra-leg) couplings, are summarized in 
Table~\ref{table:decoupled_legs}. The quantum numbers (and degeneracies) obtained
numerically are in perfect agreement with those obtained from the above 
analytical analysis.

\begin{figure}[t]
\begin{center}
	\includegraphics[width=\columnwidth]{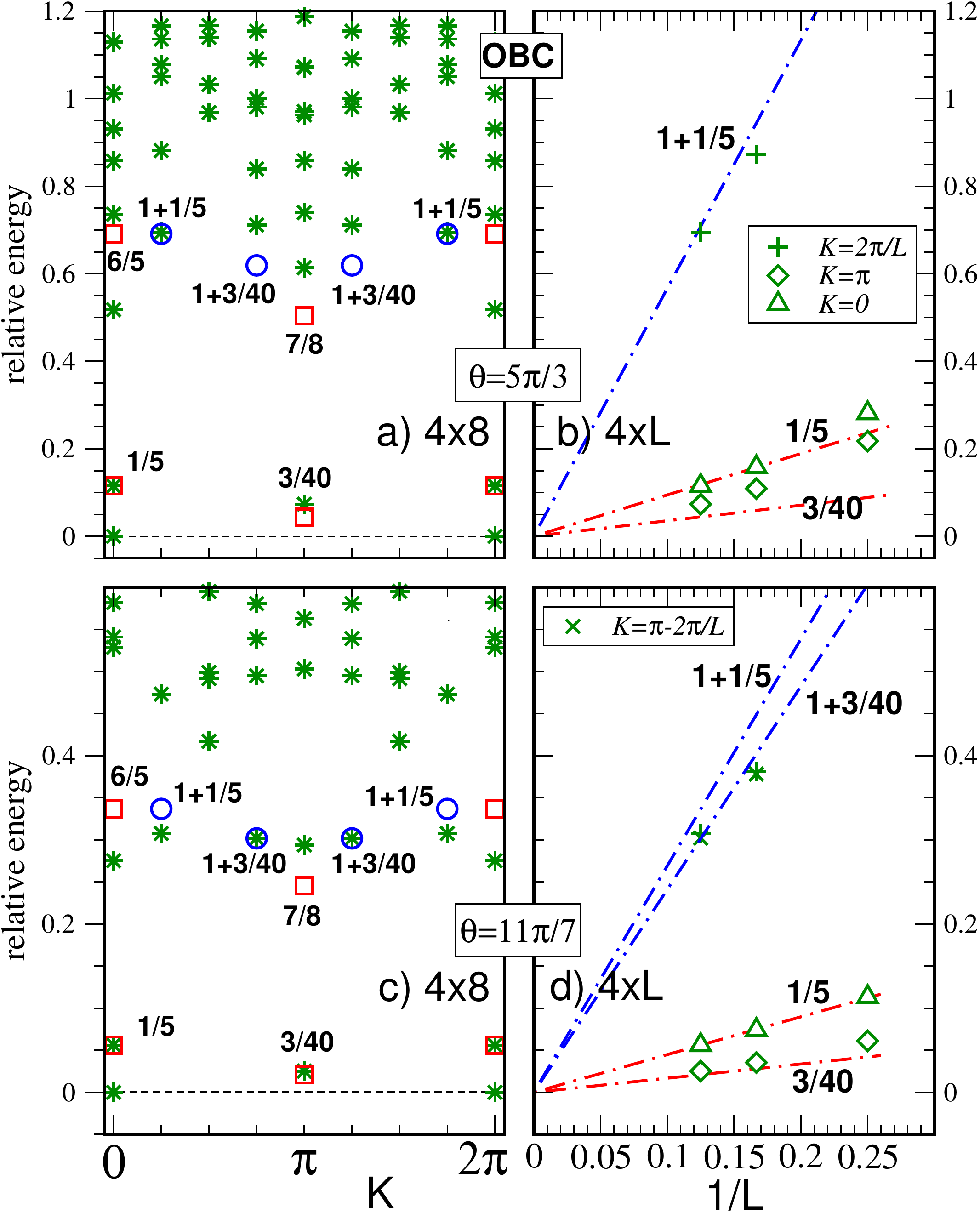}
\end{center}
\caption{
   (color online) 
   Low-energy excitations of 4-leg ladders with open boundary conditions.  
The
ground-state
energy (K=0) is used as energy reference.
   Ferromagnetic (antiferromagnetic) rung (leg) coupling are considered  i.e. giving rise to the $c=7/10$ phase of Fig.~\ref{Fig:Phasediags}(a):
   moderately strong rung coupling, $|J_{\rm rung}|/J_{\rm leg}|\sim 1.73$ ($\theta=5\pi/3$), and strong rung coupling, $|J_{\rm rung}|/J_{\rm leg}\sim 4.38$ ($\theta=11\pi/7$), are shown.   
   (a,c) Low-energy spectra of a $4\times 8$ ladder versus momentum along the ladder.
   (b,d) Finite size scalings of the low-energy levels for the two cases shown in a) and c).    Fits to c=7/10 CFT invariant spectra are provided: expected levels corresponding to primary (secondary) fields 
   of the CFT are 
   shown by red boxes (blue circles). The overall energy scale of the CFT is set by adjusting the position of
   the $K=0$, $2h=1/5$ state to the corresponding energy level of the $4\times 8$ cluster.
 }
\label{Fig:rung4_open}
\end{figure}

\section{Effects of boundary conditions}  
\label{Sec:Boundaries}

\begin{figure}
\begin{center}
	\includegraphics[width=\columnwidth]{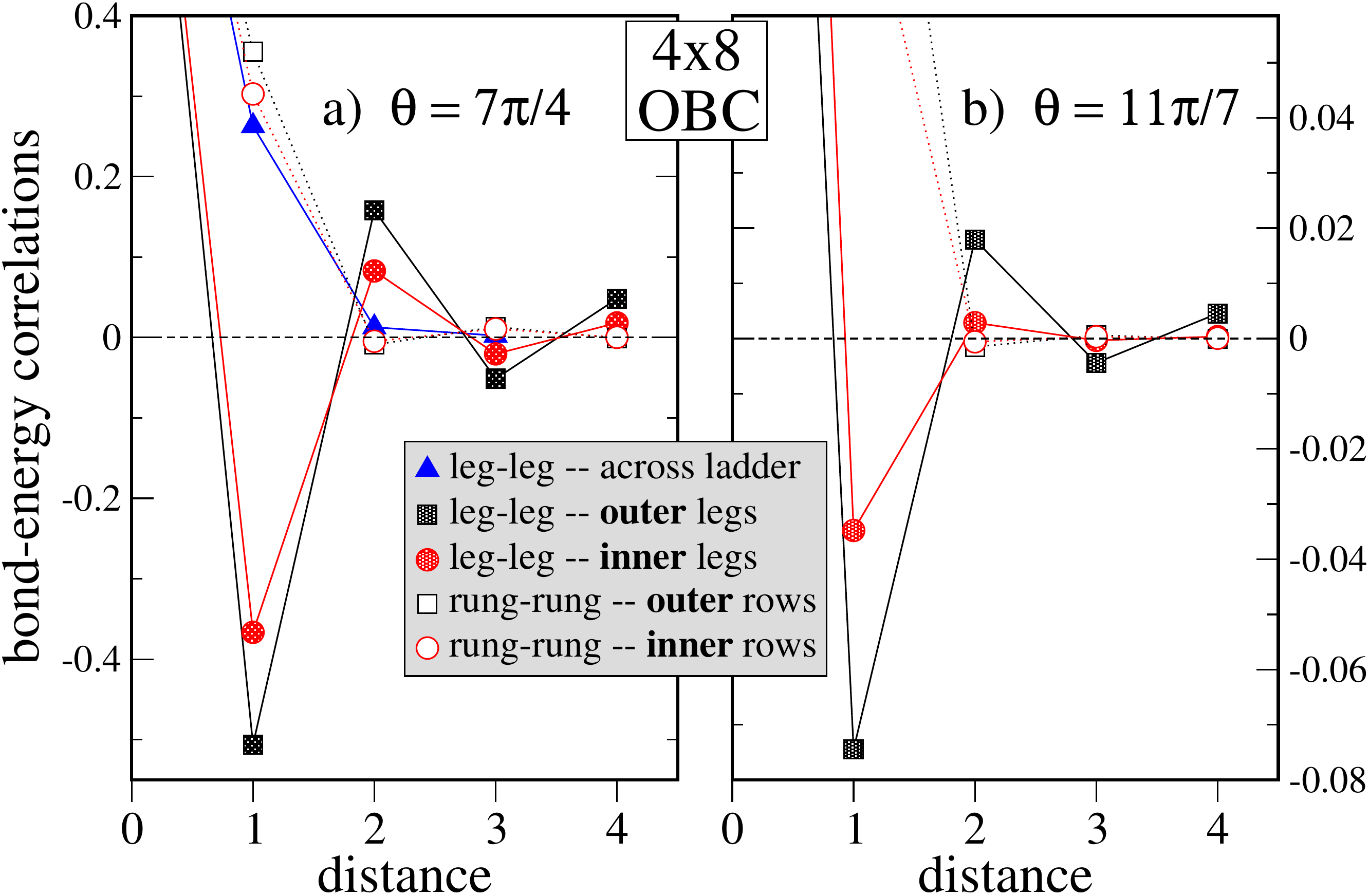}
\end{center}
\caption{
   (color online) 
   Bond-energy correlations in a $4\times 8$ ladder with open boundary conditions (OBC) 
   along the rungs (i.e. with two inner and two outer 8-site legs) as a function of the distance between two (parallel) bonds along 
   the leg direction or across the ladder. The bonds are oriented simultaneously along the rungs (rung-rung correlator) or along the legs (leg-leg correlator). 
   The disconnected part has been subtracted and the data are normalized w.r.t. the zero-distance auto-correlation.
  Data are shown for $J_{\rm rung}<0$ and $J_{\rm leg}>0$ i.e. in the $c=7/10$ phase of Fig.~\ref{Fig:Phasediags}(a). a) Isotropic couplings, $|J_{\rm rung}|/J_{\rm leg}=1$ ($\theta=7\pi/4$);
  b) strong rung coupling, $|J_{\rm rung}|/J_{\rm leg}\sim 4.38$ ($\theta=11\pi/7$). }
\label{Fig:Correlations}
\end{figure}

\begin{figure}
\begin{center}
	\includegraphics[width=\columnwidth]{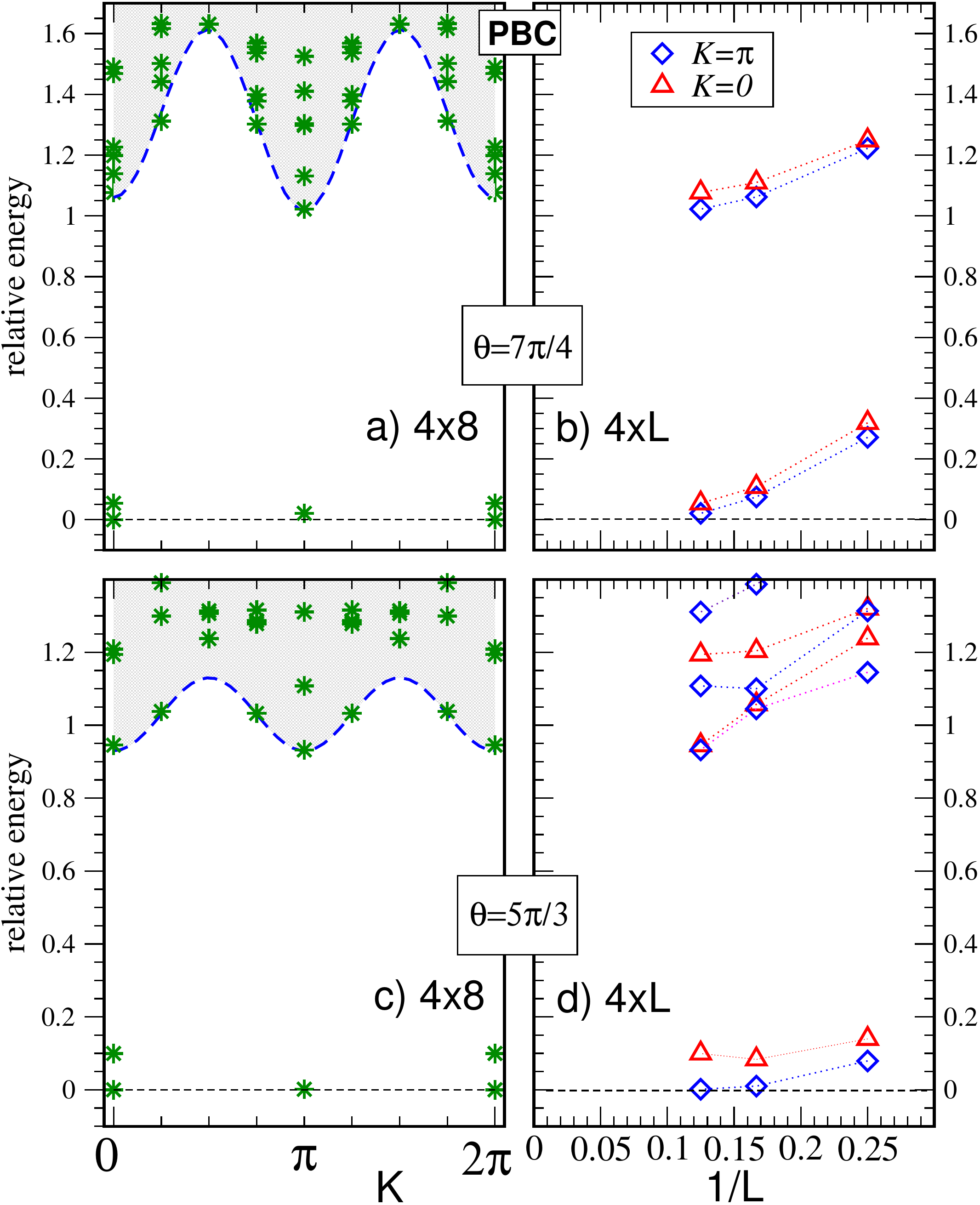}
\end{center}
\caption{
   (color online) 
   Low-energy excitations of 4-leg ladders with periodic boundary conditions.  The 
ground state
 energy (K=0) is used as energy reference.
      (a,c) Spectra of a $4\times 8$ ladder for values of $\theta$ corresponding to ferromagnetic (antiferromagnetic) rung (leg) coupling.
   Isotropic couplings, $|J_{\rm rung}|/J_{\rm leg}=1$ ($\theta=7\pi/4$), and moderately strong rung coupling, $|J_{\rm rung}|/J_{\rm leg}\sim 1.73$ ($\theta=5\pi/3$), are shown.
   (b,d) Finite size scalings of the low-energy levels for the two cases shown in a) and c) revealing 3-fold degenerate ground states and a finite gap.   
 }
\label{Fig:rung4_pbc}
\end{figure}

\begin{figure}
\begin{center}
	\includegraphics[width=\columnwidth]{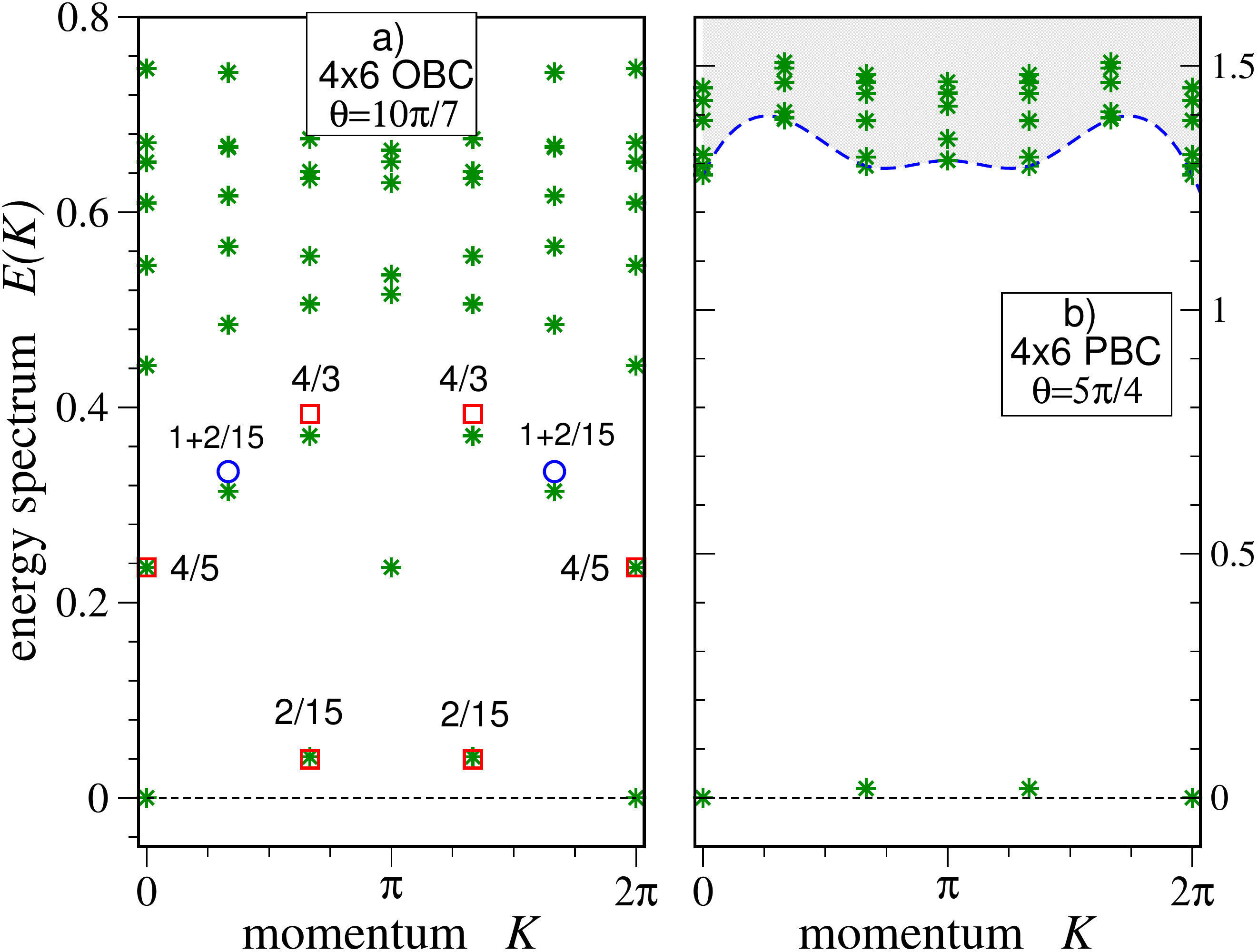}
\end{center}
\caption{
   (color online) 
   Low-energy excitations of $4\times 6$ ladders with ferromagnetic rung and leg couplings. The 
ground state
 energy (K=0) is used as energy reference.
   (a) Open boundary condition (in the rung direction). The data are obtained for strong rung coupling, $J_{\rm rung}/J_{\rm leg}\sim 4.38$ ($\theta=10\pi/7$).
   A fit to a c=4/5 CFT invariant spectrum is shown: expected levels corresponding to primary (secondary) fields of the CFT are represented by red boxes (blue circles) and the overall energy scale is set by adjusting the position of
   the K=0, 2h=4/5 energy level.
 (b) Periodic boundary condition (in the rung direction). The data are obtained for isotropic couplings, $J_{\rm rung}/J_{\rm leg}=1$ ($\theta=5\pi/4$). A large gap is seen above three quasi-degenerate 
 levels.
 }
\label{Fig:rung4_4x6}
\end{figure}

A characteristic feature of a topological phase is that it is sensitive to the topology of the underlying manifold \cite{Wen89}, which is reflected in a non-trivial ground-state degeneracy and the occurrence of gapless edge modes for open boundaries. In this Section, we will investigate the sensitivity of the anyon ladder systems to these latter effects of changing boundary conditions.
So far we considered anyon ladder systems with open boundary conditions along the rung direction 
and periodic boundary conditions along the leg directions, resulting in the topology of an annulus. Following the arguments in Refs.~\onlinecite{Gils09,Ludwig10} we interpret the observation of gapless states in the energy spectrum as the appearance of gapless {\sl edge modes} at the open boundary conditions. As a consequence, we expect the energy spectrum to gap out as we remove the gapless edge states by gluing together the open boundaries of the annulus to yield a torus geometry. 
As detailed in Section \ref{periodic} this topology change is accomplished by adding a rung coupling between the two outer legs of the ladder (and introducing the correct Dehn twist).

As an example system we consider a four-leg ladder with ferromagnetic rung coupling 
$J_{\rm rung}<0$. 
In the case of open boundary conditions along the rung direction, we find clear signatures
for gapless edge modes at these open boundaries:
First, the energy spectrum is gapless in the thermodynamic limit as shown in 
Figs.~\ref{Fig:rung4_open}. The energy eigenvalues again agree well with the expected 
conformal weights, both in the strong and intermediate coupling regimes shown in Figs.~\ref{Fig:rung4_open}c) and Figs.~\ref{Fig:rung4_open}a), respectively.
Second, correlations of the bond-energy operator decrease significantly {\sl slower} 
between two bonds located on the two {\sl outer} legs than on the {\sl inner} legs as shown in 
Fig.~\ref{Fig:Correlations}. 
In addition, the bond-bond energy correlations on the {\sl rungs} in both the two outer rows and the inner row 
of the 4-leg ladder decay more rapidly that their leg counterparts. 
We interpret these differences as evidence for a gapless edge mode
being located at the open boundaries of the ladder system and the presence of a gap in the bulk.

The occurrence of the bulk gap of this anyon ladder system becomes even more evident
when we consider the energy spectrum as we close the open boundary conditions, 
thereby removing the gapless (edge) modes. 
In Fig.~\ref{Fig:rung4_pbc} we show such clearly gapped energy spectra for periodic boundary conditions 
(in the intermediate coupling regime $\theta=5\pi/3$ and $\theta=7\pi/4$).
This observation should be contrasted with our results  for open boundary conditions and the same coupling parameters: as shown in Figs.~\ref{Fig:rung4_open}a) and \ref{Fig:rung4_open}b), for the same coupling parameter $\theta=5\pi/3$ the energy spectrum in the case of open boundary conditions nicely matches the gapless spectrum of a conformal field theory.

Furthermore, the gapped energy spectrum for periodic boundary conditions, as illustrated in
Fig.~\ref{Fig:rung4_pbc}, also reveals 
the occurrence of an unusual, non-trivial ground-state degeneracy for the anyonic
ladder system. 
For example, in the case of ferromagnetic rung coupling $J_{\rm rung}<0$ and antiferromagnetic 
leg coupling $J_{\rm leg}>0$, we observe three ground states 
\footnote{
By changing the initial conditions of the Lanczos exact diagonalization procedure, we have checked 
that each of these levels corresponds indeed to a {\sl single} energy eigenstate.
}
(one at momentum $0$ and two at momentum $\pi$) separated from the rest of the energy spectrum by a gap of order $O(1)$ in the exchange coupling strength, which become degenerate 
in the thermodynamic limit. Evidence for the latter is provided in the finite-size scaling plots of Fig.~\ref{Fig:rung4_pbc}b) and d).
It is important to notice that such a ground state degeneracy is {\sl not} due to a spontaneous dimerization along the ladder direction 
(or to any other spontaneous translation symmetry breaking). Indeed, in the case of a spontaneous dimerization, the expected ground state 
degeneracy would be a multiple of two (depending on whether the system breaks translational invariance along both ladder directions) instead of three for the anyon ladder. 

\begin{figure}
\begin{center}
	\includegraphics[width=\columnwidth]{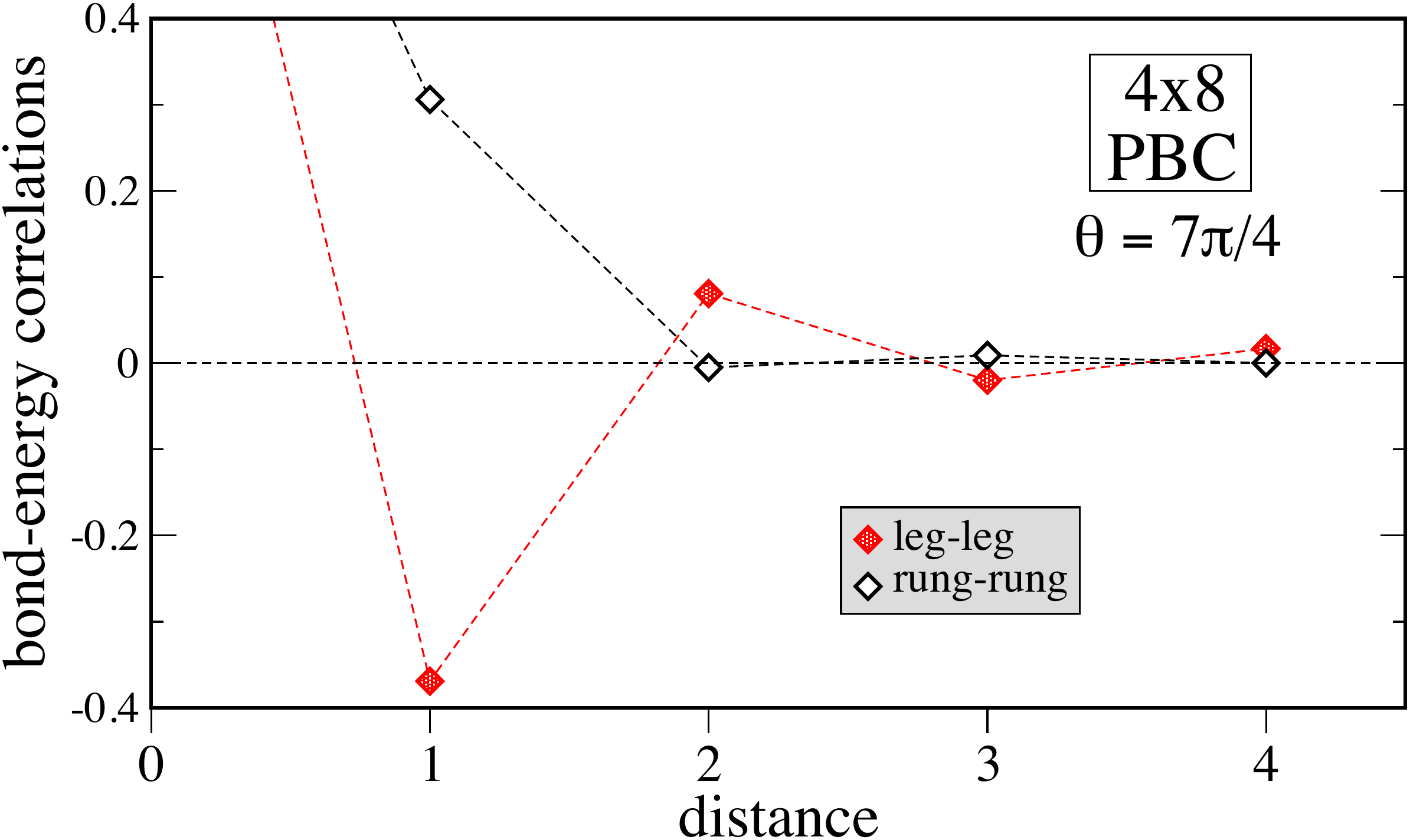}
\end{center}
\caption{
   (color online) 
   Bond-energy correlations in a $4\times 8$ ladder with {\sl periodic} boundary conditions (PBC) 
   along the rungs as a function of the distance (in the leg direction) between two (parallel) bonds.
   The bonds are oriented simultaneously along the rungs (rung-rung correlator) or along the legs (leg-leg correlator). 
   The disconnected part has been subtracted and the data are normalized w.r.t. the zero-distance auto-correlation.
  Data are shown for $J_{\rm rung}<0$ and $J_{\rm leg}>0$ i.e. in the $c=7/10$ phase of Fig.~\ref{Fig:Phasediags}(a)
  and for isotropic couplings, $|J_{\rm rung}|/J_{\rm leg}=1$ ($\theta=7\pi/4$).
  }
\label{Fig:Correlations_pbc}
\end{figure}

Further evidence for a uniform anyon ground state is provided by inspection of the correlations of the energy (rung or leg) bond operators
shown in Fig.~\ref{Fig:Correlations_pbc} for the same periodic anyon ladder. While for a dimerized system (period-2) oscillations of 
these correlations survive at arbitrarily large separations between the bonds (the amplitude is the square of the order parameter for infinite separation), our data
show in contrast a rapid vanishing of those oscillations with distance. 
 
Similarly, we have also checked that the spectrum of a $4\times 6$  ladder with both ferromagnetic rung and leg couplings, $J_{\rm rung}<0$ 
and $J_{\rm leg}<0$, 
is fully consistent with (i) a  $c=4/5$ CFT invariant spectrum when {\sl} open boundaries are used, most evidently seen for strong rung coupling in Fig.~\ref{Fig:rung4_4x6}~a)
 and (ii) a gapped spectrum and a three-fold degenerate ground state with momenta $0$ and $\pm 2\pi/3$ is found
 when using {\sl periodic} boundary conditions (i.e. removing the edges), most evidently seen for isotropic couplings in Fig.~\ref{Fig:rung4_4x6}~b). 
 Again, this degeneracy is {\sl not} connected to translation symmetry breaking but rather is a signature of
 a new (uniform) topological liquid. 
\footnote{
This behavior of the anyon ladder systems should be contrasted to the case of conventional SU(2) spin ladders.
First, SU(2) spin ladders with an even number of legs always exhibit a gap (apart for the case of simultaneous ferromagnetic rung and leg couplings for which the ground state 
is a trivial fully polarized ferromagnet). Secondly, ladders with
an odd number of legs are always gapless if open boundary conditions are used along the rung direction. When periodic boundary conditions along the rung direction are used to
form so-called spin-tubes~\cite{spintube} with an odd number of legs, dimerization in the leg direction generically sets in if the rung exchange coupling is antiferromagnetic. 
}

\section{su(2)$_k$ generalizations} 
\label{Sec:SU2k}

We now turn to the question of how the characteristic features of the $W$-leg ladder models found
for the Fibonacci theory su(2)$_3$ are generalized when considering su(2)$_k$ theories with $k>3$.
All these theories allow to define ladder models built out of generalized angular momenta $j=1/2$,
similar to the description given in Sec.~\ref{Sec:Model}. For these more general theories we can identify angular momentum $j$ with angular momentum $k/2-j$ with the highest possible allowed angular momentum thus becoming $(k-1)/4$ when $k$ is odd. 

Following the same route as taken for the Fibonacci theory, we can access most features of their
respective phase diagrams by considering the strong rung-coupling limit as presented in Sec.~\ref{Sec:StrongCoupling}. 
In particular, such an approach reveals the appearance of gapped and gapless phases  as a function of the ladder width $W$ and the level $k$. For antiferromagnetic rung-coupling $J_{\rm rung} > 0$ we find
that the odd/even effect of su(2)$_3$ occurs for all level $k$. On the other hand, for ferromagnetic rung-coupling $J_{\rm rung} < 0$ a more refined picture emerges: If the ladder width $W$ is a multiple of the level $k$, i.e. $W = 0 \mod k$, the total angular momentum on a rung is $j=0$ and we find gapped phases around this strong rung-coupling limit. If the ladder width $W$ is not a multiple of the level $k$, i.e. $W \neq 0 \mod k$, then we still expect gapped phases if the total angular momentum on a rung is an integer (thus giving rise to generalized Haldane phases \cite{Gils09}). Similarly, we expect that gapless phases are found for a total angular momentum on a rung becoming a half-integer (and $W$ not a multiple of $k$). These results are summarized in Table \ref{table:single_rung_ferro}.

This scenario also matches nicely the well-known behavior of ordinary SU(2) ladder models, which we recover when taking the limit of $k \to \infty$ for the anyonic theories.

\begin{table}[t]
\begin{center}
 \begin{tabular}{@{} cccccccccccccccccc @{}}
   \toprule
   k\; \textbackslash \;W  & 1 && 2 && 3 && 4 && 5 && 6 && 7 && 8 && 9 \\ 
   \hline
   3 & $1/2$ &$\Vert$ & $1/2$ && {\bf 0} && $1/2$ && $1/2$ && {\bf 0} && $1/2$ && $1/2$ && {\bf 0} \\ 
  5 & $1/2$ && \fbox{1} &$\Vert$ &\fbox{1} && $1/2$ && {\bf 0} && $1/2$ && \fbox{1} && \fbox{1}  && $1/2$ \\ 
  7& 1/2 && \fbox{1} && 3/2 &$\Vert$ & 3/2 &&  \fbox{1} && 1/2 && {\bf 0} && 1/2 && \fbox{1} \\ 
   9 & 1/2 &&  \fbox{1} && 3/2 &&  \fbox{2} &$\Vert$ & \fbox{2} && 3/2 && \fbox{1} && 1/2 && {\bf 0} \\ 
$\vdots$ &  $\hdots$&&&&&&&&&  &  &  & &  &   & &   \\ 
   $\infty$ & $1/2$ && \fbox{1} && 3/2 && \fbox{2} && 5/2&& \fbox{3} &&  7/2 && \fbox{4} && 9/2 \\ 
   \toprule
 \end{tabular}
 \caption{GS angular momentum (i.e. total spin) of a single {\it ferromagnetic} su(2)$_k$ rung as a function of its length $W$ (first line) and $k$ (first column). The last line gives the GS total spin of ferromagnetic Heisenberg SU(2) open chains of same lengths. When small $J_{\rm leg}$ is switched on, the corresponding $W$-leg ladders can be mapped onto single gapless effective chains (at low energies) except 
for the cases (i) marked by boxes (Haldane effective chains) or (ii) marked by "$\bf 0$" (i.e. when the GS of a single rung is a singlet). The vertical bars $\Vert$ indicate the range $W<k/2$ (see text).}
 \label{table:single_rung_ferro}
 \end{center}\end{table}


\section{Approaching the 2D limit}
\label{Sec:2D}

We conclude with a perspective on how to connect the results obtained here for $W$-leg ladders
to the thermodynamic limit of two-dimensional lattice configurations of non-Abelian anyons. 
The strong rung-coupling limit, which was useful to discuss the phases of $W$-leg ladders, turns out
to be of little help in understanding this 2D limit. The reason is that the gap of an isolated rungs vanishes as $1/W$ with increasing width, which restricts the applicability of the perturbative argument around the strong rung-couling limit to a regime of couplings $J_{\rm leg} / J_{\rm rung} < {\cal O}(1/W)$, which also vanishes as $W \to \infty$.

Instead we consider the following general symmetry argument:
In contrast to their ordinary SU(2) counterpart, the su(2)$_k$ anyonic theories lack a built-in 
continuous symmetry. In the assumed absence of an emergent continuous symmetry this reduces their ability  to undergo a spontaneous symmetry breaking transition -- such as, in two dimensions, the formation of a N\'eel state and its gapless Goldstone mode for ordinary SU(2) quantum magnets. 
Therefore one is naturally led to expect gapped quantum ground states, such as topological quantum liquids, in these anyonic systems. 
This raises the question of how these two seemingly disjunct scenarios for SU(2) and su(2)$_k$ can be reconciled when taking the $k \to \infty$ limit of the anyonic theories.
Noting that the deformation of SU(2) used to describe the anyonic systems explicitly breaks time reversal symmetry we can think of $1/k$ as the strength of a symmetry breaking field. As such we
expect the bulk gap of the 2D anyonic quantum ground state to close as one approaches the SU(2)
limit, thereby smoothly connecting the topological quantum liquids to the  N\'eel state.

The formation of a gapped bulk liquid in the thermodynamic limit is further backed by the `liquids
picture' presented in Ref.~\onlinecite{Ludwig10}. There we have argued that the interactions between
a set of non-Abelian anyons arranged on a two-dimensional lattice gives rise to the nucleation of a
new bulk-gapped (i.e.~topological) quantum liquid within the `parent liquid' of which the anyons are 
excitations of. At the spatial interface between these two distinct, bulk-gapped phases gapless edge
modes will form whose precise character can be identified from the gapless modes of one-dimensional chains of anyons \cite{Gils09}, which in turn allows for an identification of the newly formed two-dimensional bulk-gapped liquid \cite{Ludwig10}.
For the case that both the rung and leg couplings are `antiferromagnetic', i.e.~$J_{\rm rung}>0$
and $J_{\rm leg}>0$, this liquid is described by a su(2)$_{k-1}\times$su(2)$_1$ Chern-Simons theory. 
On the other hand, if both couplings are `ferromagnetic',  i.e.~$J_{\rm rung}<0$
and $J_{\rm leg}<0$, then this liquid is described a $U(1)$ Chern-Simons theory.
The case of mixed coupling signs remains open.

Anyonic generalizations of quantum magnets in the spirit of the work presented here can discussed in analogous fashion for other anyonic theories (tensor categories) and for other two-dimensional lattice geometries and interactions. We expect this to be a fruitful and broad field of research at the interface of quantum magnetism and topological states of matter.


\section*{Acknowledgments}
We thank Z. Wang for insightful discussions.
D.P.  was supported by the French National Research Agency (ANR), A.W.W.L., in part, by NSF DMR-0706140, and M.T. by the Swiss  National Science Foundation. 
We acknowledge hospitality of the Aspen Center for Physics, the Max-Planck Institute for the Physics of Complex Systems, Dresden
and the Kavli Institute for Theoretical Physics supported by NSF PHY-0551164.


\end{document}